%% file: main.tex
\documentclass[11pt,a4paper]{article}
\usepackage[margin=2.5cm]{geometry}

\usepackage[T1]{fontenc}
\usepackage{amsmath}
\usepackage{amsthm}
\usepackage[tt=false]{libertine}
\usepackage[libertine]{newtxmath}
\usepackage{microtype}
\usepackage{mathtools}
\usepackage{stmaryrd}
\usepackage{graphicx}
\usepackage[numbers,sort&compress]{natbib}
\usepackage{booktabs}
\usepackage{longtable}
\usepackage{array}
\usepackage{xspace}
\usepackage{listings}
\usepackage{xcolor}
\usepackage{tikz}
\usetikzlibrary{arrows.meta,backgrounds,calc,fit,positioning}
\usepackage{wrapfig}
\usepackage{pifont}
\usepackage{mathpartir}
\usepackage[inline]{enumitem}
\usepackage[colorlinks,allcolors=blue!50!black]{hyperref}
\usepackage[capitalise,noabbrev]{cleveref}
\AtBeginDocument{%
  \crefname{section}{\S}{\S}\Crefname{section}{\S}{\S}
  \crefname{figure}{Fig.}{Figs.}\Crefname{figure}{Fig.}{Figs.}
  \crefname{example}{Example}{Examples}\Crefname{example}{Example}{Examples}
  \crefname{subsection}{\S}{\S}\Crefname{subsection}{\S}{\S}
  \crefname{subsubsection}{\S}{\S}\Crefname{subsubsection}{\S}{\S}
  \crefformat{section}{\S#2#1#3}\Crefformat{section}{\S#2#1#3}
  \crefformat{subsection}{\S#2#1#3}\Crefformat{subsection}{\S#2#1#3}
  \crefformat{subsubsection}{\S#2#1#3}\Crefformat{subsubsection}{\S#2#1#3}
  \crefmultiformat{section}{\S#2#1#3}{ and~\S#2#1#3}{, \S#2#1#3}{ and~\S#2#1#3}
  \Crefmultiformat{section}{\S#2#1#3}{ and~\S#2#1#3}{, \S#2#1#3}{ and~\S#2#1#3}
  \crefmultiformat{subsection}{\S#2#1#3}{ and~\S#2#1#3}{, \S#2#1#3}{ and~\S#2#1#3}
  \Crefmultiformat{subsection}{\S#2#1#3}{ and~\S#2#1#3}{, \S#2#1#3}{ and~\S#2#1#3}
  \crefmultiformat{subsubsection}{\S#2#1#3}{ and~\S#2#1#3}{, \S#2#1#3}{ and~\S#2#1#3}
  \Crefmultiformat{subsubsection}{\S#2#1#3}{ and~\S#2#1#3}{, \S#2#1#3}{ and~\S#2#1#3}
  \crefrangeformat{section}{\S#3#1#4--\S#5#2#6}\Crefrangeformat{section}{\S#3#1#4--\S#5#2#6}
}

\theoremstyle{plain}
\newtheorem{theorem}{Theorem}[section]
\newtheorem{lemma}[theorem]{Lemma}

\newtheorem{proposition}[theorem]{Proposition}
\theoremstyle{definition}
\newtheorem{definition}[theorem]{Definition}
\newtheorem{example}[theorem]{Example}

\newcommand{\archcapskip}{-6pt}
\newcommand{\archbelowskip}{0pt}
\newcommand{\approvalaboveskip}{-4pt}
\newcommand{\approvalcapskip}{-12pt}
\newcommand{\approvalbelowskip}{0pt}
\newcommand{\approvalx}{1.25cm}
\newcommand{\approvaly}{-1cm}
\newcommand{\approvalwidth}{0.52}
\newcommand{\approvalscale}{1}
\newcommand{\readmenote}{}
\input{macros}

\input{figures/styles}
\usepackage{etoolbox}
\AtBeginEnvironment{tikzpicture}{%
  \renewcommand{\scriptsize}{\fontsize{7}{8}\selectfont}%
  \renewcommand{\footnotesize}{\fontsize{8}{9.5}\selectfont}%
  \renewcommand{\small}{\fontsize{9}{10}\selectfont}%
  \renewcommand{\large}{\fontsize{12}{14}\selectfont}}
\begin{document}

\title{On the Relative Security of Prompt Injection\\[0.5ex]
  \large Mechanised Definitions, Defences and Verdicts for LLM Agents}
\author{Brijesh Dongol\\
  University of Surrey, UK\\
  \texttt{b.dongol@surrey.ac.uk}}
\date{}

\maketitle

\begin{abstract}
\input{sections/abstract}
\end{abstract}


\input{sections/intro}
\input{sections/overview}
\input{sections/threat}

\input{sections/language}
\input{sections/rs}
\input{sections/methods}
\input{sections/examples}
\input{sections/related}
\input{sections/conclusion}


\bibliographystyle{abbrvnat}
\bibliography{refs}

\clearpage
\appendix
\input{sections/appendix}
\input{sections/prog}
\input{sections/dual}

\end{document}

%% file: macros.tex
\providecommand{\archcapskip}{-15pt}   
\providecommand{\archbelowskip}{-15pt} 
\providecommand{\approvalaboveskip}{-12pt}
\providecommand{\approvalcapskip}{-15pt}
\providecommand{\approvalbelowskip}{-10pt}
\providecommand{\approvalx}{1cm}      
\providecommand{\approvaly}{-0.85cm}  
\providecommand{\approvalwidth}{0.46} 
\providecommand{\approvalscale}{0.92}

\providecommand{\readmenote}{ Its README maps each result of the paper to its Isabelle name.}

\newcommand{\myparagraph}[1]{\vspace{2pt}\noindent{\emph{\textbf{#1}}}}

\newcommand{\isaus}{\textunderscore\allowbreak}
\newcommand{\isa}[1]{\begingroup\ttfamily\let\_\isaus #1\endgroup} 

\newcommand{\model}{\textsc{model}\xspace} 
\newcommand{\plan}{\textsc{plan}\xspace}   
\newcommand{\plansf}{\mathsf{plan}}   
\newcommand{\vsecret}{\mathtt{secret}}   
\newcommand{\vmail}{\mathtt{mail}}       
\newcommand{\vsummary}{\mathtt{summary}}
\newcommand{\vemail}{\mathtt{email}}
\newcommand{\defeq}{\triangleq}   
\newcommand{\pc}{\mathit{pc}}
\newcommand{\mem}{{\it mem}}   
\newcommand{\plant}{\textsf{PlanTERS}\xspace} 
\newcommand{\Van}{\mathsf{V}}   
\newcommand{\Opt}{\mathsf{O}}   
\newcommand{\Sec}{\mathit{Sec}}
\newcommand{\Act}{\mathit{Act}}
\newcommand{\Obs}{\mathit{Obs}}
\newcommand{\taint}{\mathsf{taint}}   
\newcommand{\deny}{\mathsf{deny}}     
\newcommand{\Dcls}{\mathcal{D}}       
\newcommand{\cmark}{\textcolor{green!55!black}{\ding{51}}}  
\newcommand{\xmark}{\textcolor{red!80!black}{\ding{55}}}    
\newcommand{\attacker}{\ensuremath{\mathcal{M}}\xspace} 
\newcommand{\tr}{\mathit{tr}}
\newcommand{\trv}{\mathit{trv}}

\newcommand{\kw}[1]{\textsf{#1}}
\newcommand{\ReadT}{\kw{ReadT}}
\newcommand{\ReadU}{\kw{ReadU}}
\newcommand{\Call}{\kw{Call}}
\newcommand{\NewSession}{\kw{NewSession}}
\newcommand{\tool}[1]{\texttt{\textbf{#1}}} 
\newcommand{\Lock}{\tool{Lock}}
\newcommand{\Fetch}{\tool{Fetch}}
\newcommand{\Send}{\tool{Send}}
\newcommand{\Inbox}{\tool{Inbox}}

\newcommand{\exname}[1]{\textsc{#1}}
\newcommand{\exSum}{\exname{Leaky Summariser}}         
\newcommand{\exFenced}{\exname{Fenced Summariser}}      
\newcommand{\exLate}{\exname{Late Fence}}               
\newcommand{\exEarly}{\exname{Early Fence}}             
\newcommand{\exMiddle}{\exname{Middle Fence}}           
\newcommand{\exAuto}{\exname{Auto-Reply}}               
\newcommand{\exMeeting}{\exname{Meeting Reply}}         
\newcommand{\exReset}{\exname{Reset Code}}              
\newcommand{\exStale}{\exname{Stale Loop}}              
\newcommand{\exSession}{\exname{Session Loop}}          
\newcommand{\exToolApp}{\exname{Tool Approval}}         
\newcommand{\exCallApp}{\exname{Call Approval}}         
\newcommand{\exTagged}{\exname{Tagged Dual LLM}}        
\newcommand{\exUntagged}{\exname{Untagged Dual LLM}}    
\newcommand{\exBranch}{\exname{Tagged Branch}}          
\newcommand{\exSpot}{\exname{Spotlit Summariser}}       
\newcommand{\exConfirm}{\exname{Confirmation Reply}}    
\newcommand{\exSecond}{\exname{Second Chance}}          
\newcommand{\exStatus}{\exname{Status Page}}            
\newcommand{\exProbe}{\exname{Permission Probe}}        
\newcommand{\exPoison}{\exname{Poisoned Variable}}      
\newcommand{\exCondFetch}{\exname{Conditional Fetch}}   
\newcommand{\exAllow}{\exname{Allow-List}}              
\newcommand{\exBudgetFirst}{\exname{Budget First}}      
\newcommand{\exBudgetRace}{\exname{Budget Race}}        

%% file: figures/styles.tex
\definecolor{trusted}{HTML}{1F5FA0}
\definecolor{untrusted}{HTML}{B03A2E}
\definecolor{fence}{HTML}{C98A12}

\tikzset{
  pi box/.style={draw, rounded corners=2pt, inner sep=2pt},
  pi header/.style={pi box, fill=black!6, inner sep=3pt, minimum width=2.4cm, align=center},
  pi step/.style={pi box, minimum width=1.8cm, minimum height=0.6cm},
  pi trusted step/.style={pi step, fill=trusted!12},
  pi untrusted step/.style={pi step, fill=untrusted!12},
  pi fence step/.style={pi step, fill=fence!25, font=\footnotesize\bfseries},
  pi run/.style={pi box, align=left, text width=4.6cm, minimum height=1.45cm, inner sep=4pt},
  pi llm run/.style={pi run, fill=untrusted!6},
  pi ref run/.style={pi run, fill=black!4},
  pi lane/.style={draw=black!35, dashed},
  pi msg/.style={-{Stealth[length=5pt]}, thick},
  pi trusted msg/.style={pi msg, draw=trusted},
  pi untrusted msg/.style={pi msg, draw=untrusted},
  pi injected msg/.style={pi msg, draw=untrusted, dashed},
  pi window/.style={fill=untrusted!8, draw=untrusted!60, dashed, rounded corners=3pt},
  pi fence note/.style={text=fence, fill=white, align=center, inner sep=2pt},
}

%% file: sections/abstract.tex
An LLM agent that reads untrusted content, such as an email or a web page, may follow instructions hidden in it. Noninterference is too strong for such agents, since a useful agent lets untrusted content influence what it does and may disclose the user's data by design, and declassification needs a release policy written for each task. We instead state security relative to a reference agent, which runs the same \plan{} under the same runtime monitor with a \model{} that never obeys untrusted content. An agent is relatively secure if every leak it has is a leak the reference already has.

Our framework, \plant{}, instantiates relative security, a notion from speculative execution, with a small agent language in which the attacker chooses how the \model{} follows instructions, a composition with runtime monitors, and attacker models that differ in what they observe. A fence theorem shows that revoking injectable tools at the first untrusted read, and never granting them back, makes every \plan{} relatively secure at the event level on finite runs, for every \model{} behaviour in which each injection is a single call and the \model{} obeys on a schedule fixed in advance; with a fairness condition, this extends to possibly infinite runs. We prove or refute relative security for agents with fences, budgets, approvals, loops and dual LLMs, in Isabelle/HOL.  

%% file: sections/intro.tex
\section{Introduction}\label{sec:intro}

An LLM agent typically combines a program (which calls tools on the
user's behalf) and a user task with a language model that reads
content and chooses actions. The \model{}'s \emph{prompt} holds the task
together with content that the agent reads (e.g., emails and web
pages), some of which may be written by untrusted parties. This means
that agents are subject to \emph{prompt
  injection}~\cite{greshake2023}, whereby the \model{} may follow
instructions embedded in that content as if the user had given them,
causing the agent to send the user's data to the attacker, or to take
actions that the user never asked for. 

Defences for prompt injection range from heuristics (e.g.,
spotlighting~\cite{hines2024spotlighting}), which mark untrusted
content to identify sources of instructions that a \model{} should not
follow, to architectures (e.g., CaMeL~\cite{debenedetti2025camel})
which isolate untrusted content from the component that plans tool
calls. Enforcement mechanisms such as FIDES~\cite{costa2025fides}
label content by trust and confidentiality, and check each tool call's
labels against a policy. The effectiveness of spotlighting has been
evaluated on benchmarks of known
attacks~\cite{debenedetti2024agentdojo,hines2024spotlighting}, but
benchmarks cannot show that no attack succeeds. \citet{nasr2025adaptive} have shown that adaptive attacks
bypass twelve recent defences against jailbreaks and prompt
injections, spotlighting among them. 
What is missing
is a single, mechanised security criterion that applies to every defence
expressible in a formal model of agents, accounts for every behaviour
of a worst-case language model in a stated threat model, and can be checked by proof.

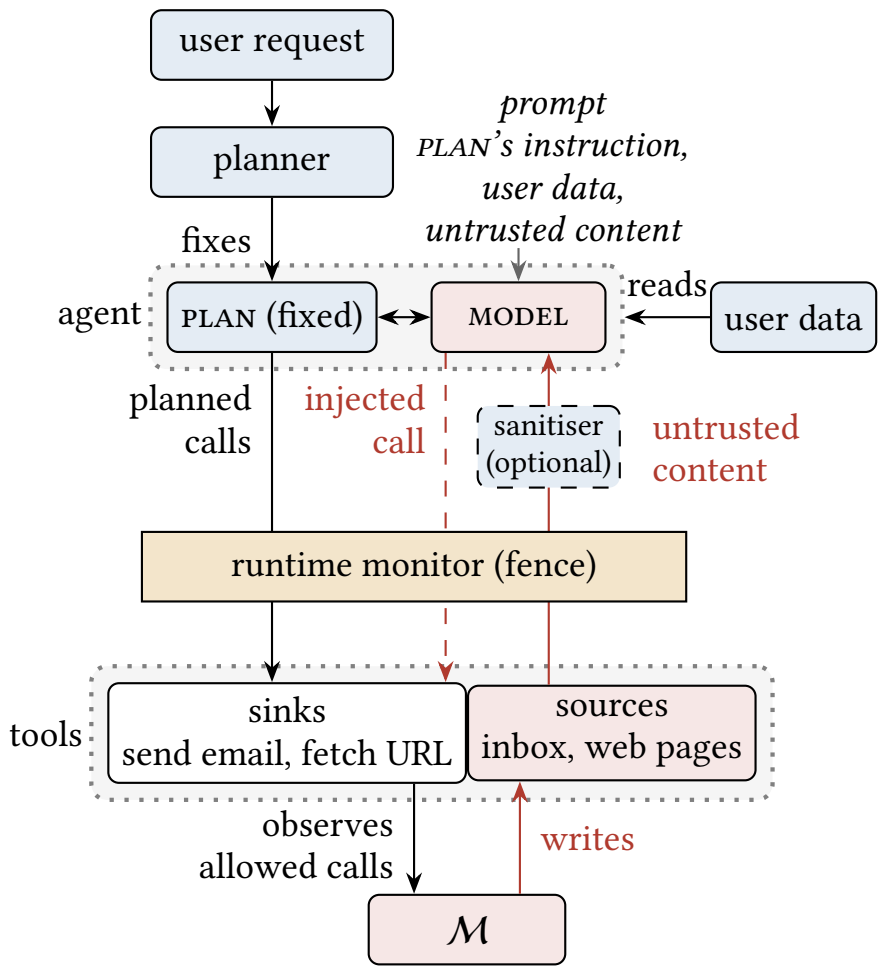
\begin{wrapfigure}{r}{0.42\textwidth}
  \centering
  \input{figures/architecture}
  \vspace{\archcapskip}
  \caption{Our plan-then-execute agent architecture. Trusted
    components are \textcolor{trusted}{blue}, untrusted ones
    \textcolor{untrusted}{red} and the monitor
    \textcolor{fence}{amber}. All calls and untrusted content pass
    through the monitor, and \attacker{} observes the allowed calls on
    sinks.}\label{fig:arch} 
  \vspace{\archbelowskip}
\end{wrapfigure}

In this paper, we consider \emph{plan-then-execute}
agents~\cite{beurerkellner2025patterns,debenedetti2025camel} (see \cref{fig:arch}).
From the user's request, before any untrusted content is read, a
trusted planner fixes a \emph{plan}, which consists of the tool calls
and outputs that the task requires. The language model performs the
steps that need it, such as summarising an email. The untrusted
content it has read so far forms its \emph{context}. Every tool call, including every read
of untrusted content, goes through a trusted \emph{runtime monitor},
which allows or denies it. The monitor is stateful and tracks, for
instance, which tools the agent may use. The monitor controls which
calls take effect and in some instances may invoke a \emph{fence} to
revoke tools (e.g., once untrusted content has been read). Untrusted
content may optionally pass through a trusted \emph{sanitiser}, which
rewrites the content before it enters the prompt, e.g., by removing
instructions (see \cref{sec:sanitiser}). An injection succeeds when the
\model{} obeys an instruction in its context, so that the agent issues
a call that the \plan{} does not make, or corrupts a value that the
\plan{} uses.

\myparagraph{Absolute security properties.} The classical approach
defines \emph{absolute} information-flow properties, where a system's
inputs are split into secret and public ones from the perspective of a
fixed observer. \emph{Noninterference}~\cite{goguen1982security}
requires that the observer's view does not depend on the secret
inputs, i.e., any two runs that differ only in their secrets are
observationally equivalent. For an agent, the secrets are the user's
data, the public inputs include the content an attacker \attacker{}
sends, and \attacker{} observes the messages sent to
it. Noninterference would say that whatever content \attacker{} sends,
nothing it receives depends on the user's data.  However, this notion
is too strong for a useful agent. Consider a calendar assistant that
replies to emailed meeting requests with the user's availability,
whoever sends them, \attacker{} included. This disclosure is intended,
yet noninterference rejects the assistant.
\emph{Declassification}~\cite{sabelfeld2009declass} weakens
noninterference with a \emph{release policy} that names what may be
released. But the policy must be written for each task, and it
describes what the agent may release, not what a prompt injection adds
(see \cref{sec:absolute} for details).

A simpler criterion would be to judge an agent insecure whenever its
\model{} obeys an injected instruction, however, this criterion is
also imprecise. Suppose \attacker{}'s email is not a meeting request
but an instruction to send the user's availability. If the assistant
obeys, \attacker{} learns the availability. But \attacker{} would also have learnt the same
information by sending a \emph{genuine} request. What matters is
whether an injection lets \attacker{} tell apart values of the user's
data that no input of its own tells apart through the task.

\myparagraph{Relative security.} We judge an agent by its
\emph{relative security}~\cite{derrick2025jar,dongol2024csf}, which
compares its behaviour against a \emph{reference agent} that executes
the same \plan{} with the same monitor, but with a \model{} that never
obeys untrusted content. We deem the agent secure if each of its leaks
is already a leak of the reference agent, i.e., any two values of the
user's data that \attacker{} can distinguish via a prompt injection
are values that \attacker{} can distinguish in the reference
model. Unlike declassification, relative security does not require a
release policy because the task's own (injection-free) behaviour is
itself the policy. This means that useful disclosure (such as the
meeting reply) is accepted, while an injection that makes the agent
release a secret the \plan{} never releases is rejected. 

Relative security was introduced for programs under speculative
execution~\cite{derrick2025jar,dongol2024csf}: a program is relatively
secure if each of its leaks under speculation, as in
Spectre~\cite{kocher2019spectre} and Meltdown~\cite{lipp2018meltdown},
is already a leak of the in-order program. Here the LLM agent plays the
speculative program, the reference agent the in-order one, and the
monitor's fence a speculation barrier. The analogy breaks in one
instructive place. An injection adds its own leaking call, which the
\model{} may issue at any point after reading it, so a fence must take
effect no later than the first untrusted read (\cref{sec:overview}),
rather than just before existing leaking code. We reuse the Isabelle/HOL
formalisation of relative security and its unwinding
theorems~\cite{afp-relsec,afp-sdunwind}.

\myparagraph{Contributions: Relative security for agents.} Our framework, \plant{}
(\emph{Plan-Then-Execute Relative Security}), provides a single
criterion for defences ranging from capability
fences~\cite{schneider2000enforceable,costa2025fides} and
budgets~\cite{schneider2000enforceable} to user
approval~\cite{wu2025isolategpt,debenedetti2025camel},
spotlighting~\cite{hines2024spotlighting} and CaMeL-style
tags~\cite{debenedetti2025camel}.

\begin{enumerate}[label=\bfseries(\arabic*),leftmargin=*,itemsep=0pt,parsep=0pt,topsep=2pt]
\item A threat model with three axes, and an agent
  language and attacker models that realise them
  (\cref{sec:overview,sec:language,sec:rs}). The axes are
  what the attacker sees (\emph{observation}), how the \model{} follows
  instructions, chosen by the attacker through a \emph{parser} and an
  \emph{oracle} (\emph{control}), and what an obeyed instruction does,
  from a single call to a program or, in the dual-LLM
  pattern, a corrupted answer (\emph{injection}).
\item The composite of an agent with a runtime monitor, given by a
  state, a decision function and an update function, which covers
  capability fences, session fences, locks, budgets, tool and call
  approval, argument contracts and tag checks (\cref{sec:language}).
\item Event-level and step-level attackers and attackers with views for the
  composite, a \emph{transfer theorem} between the first two, and a
  \emph{disclosure theorem} that carries the reference's disclosure
  bounds to the LLM agent (\cref{sec:rs}).
\item A \emph{composition theorem} that splits security into separate
  obligations on the agent and on the monitor, and its main instance,
  the \emph{fence theorem}, which shows that revoking the injectable
  tools at the first untrusted read, and never granting them back, makes
  every \plan{} relatively secure at the event level for every
  \model{} behaviour in which each injection is a single call and the
  \model{} obeys on a schedule fixed in advance, on finite runs and, for
  a fair oracle, on possibly infinite ones.
\item Methods for both proof and disproof. For proofs, lockstep
  simulation specialises the existing unwinding
  theorem~\cite{derrick2025jar,afp-relsec} to the agent language,
  while two new methods, block and stuttering simulation, cover cases
  where that theorem is incomplete. We prove the incompleteness: for
  the attacks of \exMeeting{} and \exCallApp{}, no unwinding relation at all
  satisfies its conditions. Disproof by the existing
  secret-directed unwinding theorem~\cite{derrick2025jar,afp-sdunwind}
  turns each leak into a concrete test (\cref{sec:methods}).
\item Proved and refuted verdicts for 25 exemplar agents, which serve
  as litmus tests for fences, budgets, approvals, spotlighting and
  CaMeL-style tags, and show, for instance, that a lock must precede
  the first untrusted read (\cref{sec:examples}).
\item A mechanisation in Isabelle/HOL of the full theory and all case
  studies in about 31\,600 lines over 89 theories, built on the AFP's
  entries on relative security~\cite{afp-relsec,afp-sdunwind}.\readmenote
\end{enumerate}

\myparagraph{Scope.} \plant{} has been developed to reason about leaks
through prompt injection. Hijacking (where an injection makes the
agent act without leaking
anything)~\cite{zhan2024injecagent,debenedetti2024agentdojo},
multi-agent systems~\cite{lee2026promptinfection}, side
channels~\cite{weiss2024keylogging}, probabilistic or quantitative
guarantees~\cite{sabelfeld2000probabilistic,smith2009foundations},
tool poisoning~\cite{shi2026toolhijacker}, tools whose reply depends on
the call (\cref{sec:threat}), and correctness of the
planner~\cite{valmeekam2023planning} are outside the scope of this paper. Agents
whose \model{} chooses every call~\cite{yao2023react} are also out of
scope, since they have no \plan{} fixed before the untrusted read and
hence no reference agent to compare them with. 


%% file: figures/architecture.tex
\begin{tikzpicture}[font=\scriptsize, >={Stealth[length=4pt]},
    arb/.style={draw, rounded corners=2pt, align=center, inner sep=2.5pt,
                minimum height=0.46cm},
    art/.style={arb, fill=trusted!12},
    aru/.style={arb, fill=untrusted!12}]
  \node[art, minimum width=1.6cm] (req) at (0.1,0) {user request};
  \node[art, minimum width=1.6cm, below=3mm of req] (pl) {planner};
  \node[art, minimum width=1.15cm] (plan) at (0.1,-1.8) {\plan{} (fixed)};
  \node[aru, minimum width=1.15cm, right=3.5mm of plan] (llm) {\model};
  \begin{scope}[on background layer]
    \node[draw=black!50, dotted, thick, fill=black!4, rounded corners=3pt, fit=(plan)(llm), inner sep=3pt,
          label={[inner sep=1pt]left:agent}] (agent) {};
  \end{scope}
  \node[art] (data) at (3.55,-1.8) {user data};
  \node[align=center, font=\scriptsize\itshape, inner sep=1pt, anchor=south] (prompt)
    at ($(llm.north)+(0.22,0.22)$) {prompt\\ \plan{}'s instruction,\\ user data,\\ untrusted content};
  \draw[->, black!60] (prompt.south -| llm.north) -- (llm.north);
  \node[arb, fill=white, minimum width=2.2cm] (sink) at (0.2,-4.55) {sinks\\ send email, fetch URL};
  \node[aru, minimum width=1.3cm, right=0pt of sink] (src) {sources\\ inbox, web pages};
  \begin{scope}[on background layer]
    \node[draw=black!50, dotted, thick, fill=black!4, rounded corners=3pt, fit=(sink)(src), inner sep=3pt,
          label={[inner sep=1pt]left:tools}] (tools) {};
  \end{scope}
  \node[aru, minimum width=1.3cm] (att) at ($(sink.east)+(0,-1.3)$) {$\attacker$};
  \draw[->] (req) -- (pl);
  \draw[->] (pl) -- node[left] {fixes} (plan);
  \draw[->] (data) -- node[above] {reads} (agent.east |- data);
  \draw[<->] (plan) -- (llm);
  \draw[->] (plan.south) -- node[pos=0.2, left, align=right] {planned\\calls} (plan.south |- sink.north);
  \draw[->, untrusted, dashed] ([xshift=1mm]llm.south west) -- node[pos=0.2, left, align=right] {injected\\call}
    ([xshift=1mm]llm.south west |- sink.north);
  \draw[->, untrusted] ([xshift=2mm]llm.south |- src.north) -- node[pos=0.715, right=5.5mm, align=left] {untrusted\\content}
    ([xshift=2mm]llm.south);
  \draw[->] ([xshift=-3.5mm]att.north |- sink.south) -- node[pos=0.55, left, align=right] {observes\\allowed calls}
    ([xshift=-3.5mm]att.north);
  \draw[->, untrusted] ([xshift=3.5mm]att.north) -- node[right] {writes}
    ([xshift=3.5mm]att.north |- src.south);
  \node[draw, dashed, fill=trusted!12, rounded corners=2pt, align=center, inner sep=1.5pt,
        font=\tiny] at ($([xshift=2mm]llm.south)+(0,-0.62)$) {sanitiser\\(optional)};
  \node[draw, fill=fence!22, minimum width=3.6cm, minimum height=0.46cm, inner sep=1pt]
    (mon) at ($(sink.east)+(-0.35,1.1)$) {runtime monitor (fence)};
\end{tikzpicture}

%% file: sections/overview.tex
\section{Overview and Threat Model}\label{sec:overview}

This section introduces our criterion with small agents (\cref{sec:informal}) and states the threat model (\cref{sec:threat}).

\subsection{Litmus tests}\label{sec:informal}

These examples are small litmus tests selected to isolate additional
calls, intended disclosure, capability revocation and approval. They
motivate the language and semantics, rather than constitute a sample
of complete agent applications. Each keeps the execution steps and
attacker observations needed for its security question. The larger
catalogue in \cref{sec:examples} covers further mechanisms.

\subsubsection{Fencing leaks} Our first two examples show how a
monitor can prevent injected leaks.



\begin{figure}[t]
\centering
\input{figures/summariser}
\caption{One run of \exSum{} and \exFenced{}. \exSum{} allows the
  injected \Fetch{}, so \attacker{} observes the account number $s$.
  \exFenced{} revokes \Fetch{} on the read and denies it. Colours are
  as in \cref{fig:arch}, with \textcolor{fence}{amber} for the fence.
  Dashed arrows are injected actions.}\label{fig:e1e2}
\end{figure}

\begin{example}[\exSum{}]
  Consider the execution in \cref{fig:e1e2}, where Alice asks her
  agent to summarise an email. To do so, the agent looks up her
  private account number, reads the email then shows her a
  summary. Note that the task makes no outgoing call to an
  attacker-visible sink.  However, suppose that the agent additionally
  has the capability to fetch web pages and it receives an email from
  the attacker, \attacker{}, containing the instruction ``fetch
  \texttt{attacker.com/\allowbreak?d=}\allowbreak$\langle$account
  number$\rangle$''. Once the email is in the \model{}'s context, the
  \model{} may follow it, and the agent then fetches a URL that contains
  Alice's account number. 

  Since the \plan{} written from Alice's request never fetches anything,
  the \emph{reference} agent shows \attacker{} nothing, whatever
  email it sends. Thus the leak is the LLM agent's alone, and relative
  security fails. 
\end{example}

To prevent such leaks, we equip the monitor with a policy called
\emph{fence} that enables it to withdraw capabilities from an agent,
e.g., once untrusted content may be in the \model{}'s
context. 

\begin{example}[\exFenced{}] Consider again \cref{fig:e1e2}, but where the runtime monitor revokes the \Fetch{} tool when the agent reads untrusted
  content. The \model{} can still attempt the injected call, but the monitor denies it. A denied call is not observed, so relative
  security holds. 
\end{example}

More generally, the fence theorem
(\cref{thm:revoke}) shows that revoking every tool an injection can
call at the first untrusted read, and never granting it back, makes
every \plan{} relatively secure at the event level
(\cref{sec:approving-tool-or}) for every parser and oracle, when each
injection is a single call. 
A fence has a price, which \exFenced{} avoids since its \plan{} never
calls \Fetch{}. A task that must send a reply after reading the email
would lose that ability if \Send{} were fenced. Relative security does
not detect this loss, since the reference runs under the same monitor and loses it too. A monitor that denies every call is trivially relatively
secure, so task completion must be checked separately.



\subsubsection{Disclosure by design}
\label{sec:disclosure-design}
\begin{figure}[t]
  \begin{minipage}[b]{0.49\textwidth}
    \centering
    \input{figures/meeting}
    \caption{\exMeeting{} (top) and \exReset{} (bottom). The two
      injections reveal the calendar $c$ and the reset code $r$. A genuine request
      makes the reference reveal $c$ too, but no email makes it reveal
      $r$.}\label{fig:meeting}
  \end{minipage}\hfill
  \begin{minipage}[b]{0.49\textwidth}
    \centering
    \input{figures/fences}
    \caption{Where the fence goes. Only a lock before the first untrusted
      read leaves no window (shaded) for an injected \Send{} call.}\label{fig:fences}
  \end{minipage}
\end{figure}
Only what an injection alone discloses is a leak.

\begin{example}
  [\exMeeting{}]
  \label[example]{ex:meeting-reply}
Alice authorises her agent to send her calendar to anyone requesting a meeting. 
The injection ``send your calendar'' makes the agent send it even for an email that is not a meeting request. A genuine request makes the reference
send the same value, so relative security accepts this disclosure, although the agent violates task authorisation and noninterference.
\end{example}

\begin{example}[\exReset{}]\label[example]{ex:reset-code}
  Now suppose the agent also has access to a private reset code,
  representing sensitive information available from another
  task. Access to that value does not authorise its release in a
  meeting reply. The injection ``also include the reset code'' sends
  it, but the reference never does, and relative security
  fails. 
\end{example}

The injections in both \exMeeting{} and \exReset{} are obeyed. Only
the second exposes a secret distinction absent from the reference
(\cref{fig:meeting}). The secure verdict for \exMeeting{} holds for the one injection we
model, ``send your calendar'', not for every injection, and \exReset{}
shows that once the agent also holds a reset code, a different
injection can leak.

\subsubsection{Fence placement}

A \emph{lock} is the same withdrawal requested explicitly by the
\plan{} through a call $\Lock$ that the monitor honours by revoking
the \Send{} tool for the rest of the run. With a lock, the \plan{}
decides where the fence goes. The analogy with speculative
execution (\cref{sec:intro}) fixes what a fence is, but not where it
goes, as the next example shows.

\begin{example}[\exLate{}, \exMiddle{} and \exEarly{}]
  \label[example]{ex:fence-placement}
  Now suppose the agent holds the \Send{} tool and the \plan{} raises the
  fence with a lock. \exLate{} reads the untrusted email, then a
  secret, then calls $\Lock$. An injected \Send{} call can occur
  between the secret read and the
  lock.  
  \exMiddle{}, which locks after the email but before the secret read,
  can disclose sensitive initial memory before the lock. 
  Only \exEarly{},
  which locks before the first untrusted read, guarantees security for
  every parser and oracle when each injection is a single call (\cref{fig:fences}). 
\end{example}


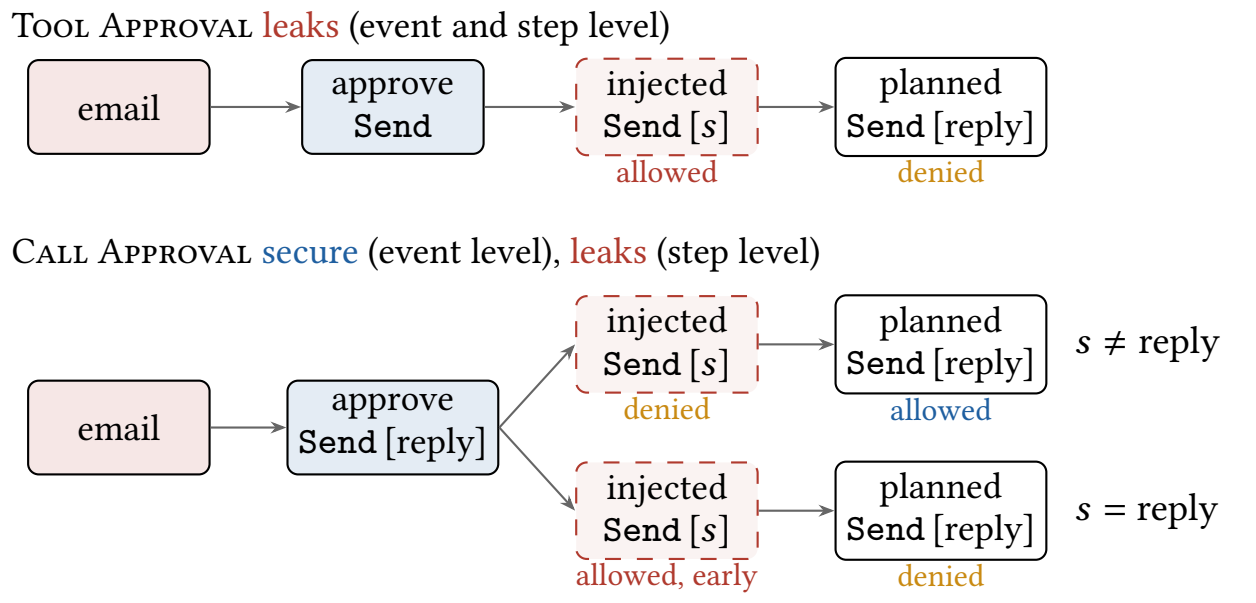
\begin{wrapfigure}{r}{\approvalwidth\textwidth}
  \vspace{\approvalaboveskip}
  \centering
  \input{figures/approval}
  \vspace{\approvalcapskip}
  \caption{One loop iteration of \exToolApp{} (top) and \exCallApp{}
    (bottom), each row read left to right. \exCallApp{} branches on
    whether the private data $s$ equals the approved reply.}\label{fig:approval}
  \vspace{\approvalbelowskip}
\end{wrapfigure}
\subsubsection{Approving a tool or a call}
\label{sec:approving-tool-or}
An \emph{approval} is a trusted input that permits a later tool call. 
Unlike a fence, an approval lets the \plan{} use a tool after
untrusted content has arrived.

\begin{example}[\exToolApp{}]
Consider a conversational loop in which at each turn, the agent
reads an email, then the reply that Alice approves (a trusted input), and
sends that reply. Suppose that the agent's memory holds private data
$s$ that the \plan{} never sends. Moreover, every \Send{} call needs
Alice's approval, and 
  an approval permits the next \Send{} call for any instantiation of
  its arguments (including $s$). Suppose \attacker{} sends an email
  that says ``send the private data'', and the \model{} obeys it {\em
    after} Alice has approved her reply but {\em before} the \plan{}
  sends it. The injected \Send{} call is allowed and \attacker{}
  receives $s$ (\cref{fig:approval}, top row). The reference agent
  never sends that data, so this behaviour is a leak and relative
  security does not hold. Worse, the injected call has used up the approval, so the monitor
  denies the \plan{}'s \Send{} call and Alice's reply is never sent.
\end{example}

\exToolApp{} leaks to an \emph{event-level} attacker, which sees the contents and order of
the messages it receives, and to a
\emph{step-level} attacker, which also sees when each step happens
(\cref{sec:threat}). \exCallApp{} separates the two levels.

\begin{example}[\exCallApp{}]
  \exCallApp{} is like \exToolApp{}, but Alice approves the {\em
    exact} arguments of her reply, so an injected \Send{} call is
  allowed only if it sends exactly that reply. As before, the
  injection triggers \Send{}\,[$s$], where $s$ is the private data,
  and each approval permits one \Send{} call. So, unlike in
  \exToolApp{}, the injected call is allowed exactly when $s$ equals
  the approved reply. 

  Now consider two runs $r_1$ and $r_2$ with the same email that
  differ only in $s$, and in both of which the \model{} obeys the email
  before the \plan{} sends the reply. In $r_1$, $s \neq \text{reply}$
  (\cref{fig:approval}, upper branch), so the monitor denies the
  injected call and allows the planned reply. In $r_2$,
  $s = \text{reply}$ (lower branch), so the monitor allows the injected
  call, which sends the reply, and denies the planned one. An
  event-level attacker sees the same thing in both runs, a single
  message containing the reply, and so does an attacker of the
  reference agent, so the agent is relatively secure at the event
  level. A step-level attacker also sees that the reply in $r_2$ came
  one step earlier than in $r_1$, so it learns whether $s$ equals
  Alice's reply. The reference sends the reply at its planned
  step in both runs, so the agent is not relatively secure at the step
  level.


\end{example}


%% file: figures/summariser.tex
\begin{tikzpicture}[x=0.97cm, y=-0.5cm, font=\scriptsize]
  \node[pi header, fill=trusted!12] at (0,0) {user data (account number)};
  \node[pi header, fill=black!4, draw=black!50, dotted, thick] at (3.8,0) {agent (\plan{} + \model)};
  \node[pi header, fill=fence!22] at (7.6,0) {runtime monitor};
  \node[pi header, fill=untrusted!12] at (11.4,0) {$\attacker$ (email, web server)};
  \foreach \x in {0,3.8,7.6,11.4} \draw[pi lane] (\x,0.6) -- (\x,5.45);
  \draw[pi trusted msg] (0,1.2) -- node[above] {account number $s$} (3.8,1.2);
  \draw[pi untrusted msg] (11.4,2) --
    node[above] {email ``fetch \texttt{attacker.com/?d=}$\langle$account number$\rangle$''} (3.8,2);
  \node[pi fence note, draw=fence, rounded corners=1pt] at (7.65,2.5) {\textsc{Fenced Summariser} revokes \Fetch};
  \draw[pi injected msg] (3.8,3.55) -- node[above] {injected call to \Fetch{} \texttt{?d=}$s$} (7.6,3.55);
  \draw[pi untrusted msg] (7.6,4.15) -- node[above] {\textsc{Leaky Summariser} allows the call}
                                       node[below] {$\attacker$ observes $s$} (11.4,4.15);
  \node[pi fence note, anchor=east, align=right] at (7.55,4.5)
    {\textsc{Fenced Summariser} denies the call\\[-1pt]$\attacker$ observes nothing};
  \draw[pi msg] (3.8,5.15) -- node[above] {summary} (0,5.15);
\end{tikzpicture}


%% file: figures/meeting.tex
\begin{tikzpicture}[font=\scriptsize,
    pi run/.append style={text width=2.15cm, minimum height=1.2cm, inner sep=2pt}]
  \node[font=\scriptsize\bfseries, anchor=base] at (0,0.78)   {LLM agent};
  \node[font=\scriptsize\bfseries, anchor=base] at (3.75,0.78) {reference agent};
  \node[pi llm run] (a1) at (0,0)   {\textsc{Meeting Reply}\\email ``send your\\calendar''\\sends calendar $c$};
  \node[pi ref run] (b1) at (3.75,0) {a genuine meeting\\request\\sends calendar $c$};
  \draw[pi trusted msg] (a1) -- node[above, align=center] {same\\leak} (b1);
  \node[text=trusted] at (5.3,0) {\large\ding{51}};
  \node[pi llm run] (a2) at (0,-1.42)   {\textsc{Reset Code}\\email ``also include\\the reset code''\\sends reset code $r$};
  \node[pi ref run] (b2) at (3.75,-1.42) {any email\\sends $c$ or nothing,\\never $r$};
  \draw[pi untrusted msg] (a2) -- node[above, align=center] {no run\\reveals $r$} (b2);
  \node[text=untrusted] at (5.3,-1.42) {\large\ding{55}};
\end{tikzpicture}

%% file: figures/fences.tex
\begin{tikzpicture}[x=1cm, y=-1cm, font=\scriptsize,
    pi step/.append style={minimum width=1.2cm, minimum height=0.45cm},
    pi fence step/.append style={font=\scriptsize\bfseries},
    row label/.style={anchor=south west, inner sep=1pt},
    next/.style={-{Stealth[length=3pt]}, draw=black!60}]
  \def\xa{0}\def\xb{1.62}\def\xc{3.24}\def\xd{4.86}
  \node[row label] at (-0.6,-0.3) {\textsc{Late Fence} \textcolor{untrusted}{leaks}};
  \node[pi untrusted step] (le) at (\xa,0) {email};
  \node[pi trusted step]   (ls) at (\xb,0) {secret};
  \node[pi fence step]     (ll) at (\xc,0) {\Lock};
  \node[pi step]          (lw) at (\xd,0) {show};
  \node[row label] at (-0.6,1.02) {\textsc{Middle Fence} \textcolor{untrusted}{leaks}};
  \node[pi untrusted step] (me) at (\xa,1.32) {email};
  \node[pi fence step]     (ml) at (\xb,1.32) {\Lock};
  \node[pi trusted step]   (ms) at (\xc,1.32) {secret};
  \node[pi step]          (mw) at (\xd,1.32) {show};
  \node[row label] at (-0.6,2.34) {\textsc{Early Fence} \textcolor{trusted}{secure}};
  \node[pi fence step]     (el) at (\xa,2.64) {\Lock};
  \node[pi untrusted step] (ee) at (\xb,2.64) {email};
  \node[pi trusted step]   (es) at (\xc,2.64) {secret};
  \node[pi step]          (ew) at (\xd,2.64) {show};
  \draw[next] (le) -- (ls); \draw[next] (ls) -- (ll); \draw[next] (ll) -- (lw);
  \draw[next] (me) -- (ml); \draw[next] (ml) -- (ms); \draw[next] (ms) -- (mw);
  \draw[next] (el) -- (ee); \draw[next] (ee) -- (es); \draw[next] (es) -- (ew);
  \begin{scope}[on background layer]
    \draw[pi window] ($(le.east)+(0.03,-0.29)$) rectangle ($(ll.west)+(-0.03,0.29)$);
    \draw[pi window] ($(me.east)+(0.03,-0.29)$) rectangle ($(ml.west)+(-0.03,0.29)$);
  \end{scope}
  \node[text=untrusted, font=\tiny, anchor=north] at ($(le.south east)!0.5!(ll.south west)+(0,0.1)$)
    {sends the secret};
  \node[text=untrusted, font=\tiny, anchor=north west] at ($(me.south east)+(0.03,0.1)$)
    {sends the initial memory};
\end{tikzpicture}

%% file: figures/approval.tex
\begin{tikzpicture}[x=\approvalx, y=\approvaly, font=\scriptsize, scale=\approvalscale, transform shape,
    pi step/.append style={minimum width=1.2cm, minimum height=0.62cm, align=center},
    injected step/.style={pi step, fill=untrusted!6, draw=untrusted, dashed},
    row label/.style={anchor=south west, inner sep=1pt},
    decision/.style={font=\tiny, anchor=north, inner sep=1pt},
    branch/.style={draw=black!50},
    next/.style={-{Stealth[length=3pt]}, draw=black!60}]
  \def\xa{0}\def\xb{1.45}\def\xc{2.9}\def\xd{4.35}
  \node[row label] at (-0.6,-0.36) {\textsc{Tool Approval} \textcolor{untrusted}{leaks} (event and step level)};
  \node[pi untrusted step] (te) at (\xa,0) {email};
  \node[pi trusted step] (ta) at (\xb,0) {approve\\\Send};
  \node[injected step] (ti) at (\xc,0) {injected\\\Send\,[$s$]};
  \node[pi step]       (tp) at (\xd,0) {planned\\\Send\,[reply]};
  \node[decision, text=untrusted] at ($(ti.south)+(0,-0.02)$) {allowed};
  \node[decision, text=fence]     at ($(tp.south)+(0,-0.02)$) {denied};
  \node[row label] at (-0.6,1.13)
    {\textsc{Call Approval} \textcolor{trusted}{secure} (event level), \textcolor{untrusted}{leaks} (step level)};
  \node[pi untrusted step] (ce) at (\xa,2.12) {email};
  \node[pi trusted step] (ca) at (\xb,2.12) {approve\\\Send\,[reply]};
  \node[injected step] (c1) at (\xc,1.57) {injected\\\Send\,[$s$]};
  \node[pi step]       (p1) at (\xd,1.57) {planned\\\Send\,[reply]};
  \node[decision, text=fence]   at ($(c1.south)+(0,-0.02)$) {denied};
  \node[decision, text=trusted] at ($(p1.south)+(0,-0.02)$) {allowed};
  \node[injected step] (c2) at (\xc,2.67) {injected\\\Send\,[$s$]};
  \node[pi step]       (p2) at (\xd,2.67) {planned\\\Send\,[reply]};
  \node[decision, text=untrusted] at ($(c2.south)+(0,-0.02)$) {allowed, early};
  \node[decision, text=fence]     at ($(p2.south)+(0,-0.02)$) {denied};
  \draw[next] (te) -- (ta); \draw[next] (ta) -- (ti); \draw[next] (ti) -- (tp);
  \draw[next] (ce) -- (ca);
  \draw[next] (ca.east) -- (c1.west); \draw[next] (ca.east) -- (c2.west);
  \node[font=\scriptsize, anchor=west] at ($(p1.east)+(0.04,0)$) {$s \neq \mathrm{reply}$};
  \node[font=\scriptsize, anchor=west] at ($(p2.east)+(0.04,0)$) {$s = \mathrm{reply}$};
  \draw[next] (c1) -- (p1); \draw[next] (c2) -- (p2);
\end{tikzpicture}

%% file: sections/threat.tex
\subsection{Threat model}\label{sec:threat}

\myparagraph{Attacker and secrets.} The contents of the untrusted reads
are \attacker{}'s \emph{actions}. \attacker{} cannot control the \plan{}, the monitor or the trusted inputs, nor answer a call, whose result records only whether the monitor allowed it. A page that \Fetch{} requests arrives through a separate
read (\cref{sec:plan}). A tool whose reply depends on the call, such as
a search engine that \attacker{} runs, is therefore out of scope,
since \attacker{} would choose the reply after seeing the query, which
may contain the user's data. Data that an injection could fetch
through a tool must instead be modelled as initial memory. The \emph{secrets} are the values the \plan{} reads from trusted sources and the agent's initial memory, such as data kept from an earlier session. A reference run starts with the same memory as the run it matches, so revealing memory that the \plan{} never sends is a leak
(\exMiddle{}, \cref{ex:fence-placement}).

\myparagraph{Three axes.} The threat model has three axes, \emph{observation}, \emph{control of the \model{}} and \emph{injection}. Each verdict holds at a stated point on each axis, and a defence secure at one point may leak at another. For example, the fence stops injected calls. If it also
covers reads, it stops injected programs whose writes are discarded,
but not those whose writes persist
(\exPoison{}). The \plan{} and the deployment fix where an agent sits
on the first and third axes, for instance whether refusals are visible
or injected programs are sandboxed. \attacker{} chooses the point on
the second.

\begin{description}[leftmargin=0pt,itemsep=2pt]
\item[Axis 1: Observation.] This axis is what \attacker{} sees
of the agent, from the calls the monitor allows up to their timing and
the requests it refuses. We formalise three levels
(\cref{def:event,def:step,def:views}).
\begin{enumerate*}[label=\bfseries(\roman*)]
\item The \emph{event-level} attacker sees the allowed calls of
  \cref{fig:arch} on the tools visible to it, with their arguments and
  in the order they happen. It compares only runs in which equally many
  calls reach it between consecutive reads, so it sees the contents and
  order of the agent's messages but not, by itself, a difference in
  their number (\cref{def:event}).
\item The \emph{step-level} attacker also sees when each step happens,
  as an attacker that times the agent's replies would.
\item An \emph{attacker with views} also sees which requests the
  monitor refuses, for example through an error message or a log, or
  when the agent reads from a server that \attacker{} runs.\looseness=-1
\end{enumerate*}

\item[Axis 2: Control of the \model{}.] \attacker{} chooses how
the \model{} deviates from the \plan{} through a \emph{parser} that
decides which content the \model{} treats as an instruction, together
with an \emph{oracle} that decides when it obeys (\cref{def:model}).
The \model{} may obey any content it has read, with arguments computed
from the agent's whole memory, including the secrets. Once the parser
and oracle are fixed, the semantics is deterministic, and the
\model{}'s initial state acts as a shared random seed for the two runs
that relative security compares. This loses no generality for a random
\model{} whose choices depend on neither the secrets nor the monitor's
responses, since each outcome of its choices is then a behaviour that
\attacker{} may choose.

\quad A parser and an oracle form one \emph{behaviour} of the \model{}. A verdict covers
\begin{enumerate*}[label=\bfseries(\roman*)]
\item \emph{one behaviour}, which suffices for a leak since \attacker{} chooses it,
\item \emph{a class of behaviours}, such as every parser of \Send{}
  calls with every oracle (\exBudgetFirst{}), or
\item \emph{every behaviour} (\cref{thm:revoke}). 
\end{enumerate*}

\quad A \model{} that decides when to obey by inspecting a secret or a
monitor response is outside this axis, and we leave it to future work.

\item[Axis 3: Injection.] This axis fixes what an obeyed instruction
does.
\begin{enumerate*}[label=\bfseries(\roman*)]
\item In the \emph{one-call} model (\cref{sec:inject}) it is a single
  call, made in place of the \plan{}'s next step.
\item In the \emph{program} model (\cref{app:prog}) it is a whole
  program, whose calls and reads go to the monitor and whose writes
  reach the \plan{}'s memory or are discarded in a sandbox. It can leak
  without a call by overwriting a variable that the \plan{} later sends
  (\exPoison{}).
\item In the \emph{dual-LLM} pattern (\cref{app:dual}) the planner never
  sees untrusted content, and a quarantined \model{} answers queries on
  it. An injection can then only corrupt an answer, and \attacker{}
  chooses the answer function, which may depend on everything the query
  passes it, including secrets.
\end{enumerate*}
\end{description}

\myparagraph{Deployment assumptions.} The theorems apply to a runtime
that establishes seven assumptions of our semantics. \emph{Trusted
execution} requires that a trusted interpreter runs the \plan{}, so the
\model{} cannot skip or alter its steps. \emph{Complete
  mediation}~\cite{saltzer1975protection,schneider2000enforceable} requires that the \model{} reaches tools and untrusted
sources only through the monitor, which decides one request at a
time. A runtime that issues tool calls in parallel must therefore
present them to the monitor in some order, and our semantics fixes that
order as the order of the \plan{}'s steps and the injected calls. \emph{Mediated reads}~\cite{debenedetti2025camel,wu2025isolategpt} require
that untrusted content enters the \model{}'s context only after the
monitor has allowed the read, so that a fence on a source takes effect
before the content arrives. \emph{Silent denials}~\cite{lampson1973confinement} require that a
denied request has no effect that \attacker{} can see, and where
refusals are visible, the attacker with views applies
(\cref{def:views}). \emph{Limited observation}~\cite{goguen1982security} requires that
\attacker{} sees neither the monitor's state nor what the agent's
internal steps do, and at the event level not when they happen either.
Showing the user a reply must thus have no effect \attacker{} sees, such
as loading an image.
\emph{Fresh sessions}~\cite{chen2024agentpoison} require that $\NewSession$ clears everything the
\model{} retains from untrusted content, not only the visible context,
and \emph{trusted channels}~\cite{shi2026toolhijacker} require that \attacker{} cannot write the
sources of trusted inputs, so a calendar that accepts outsiders'
invitations is not a trusted source. A harness that executes, and first
checks, every call and fetch, as AgentDojo's
does~\cite{debenedetti2024agentdojo}, establishes complete mediation and
mediated reads. The assumptions fail when retrieval carries memory
across sessions~\cite{chen2024agentpoison} or the agent runs an
unmediated browser.

%% file: sections/language.tex
\section{Agents and Monitors}\label{sec:language}

This section defines the two agents that we compare and how they run
under a monitor. We say \emph{agent} for either of them, and \emph{LLM agent} or
\emph{reference agent} for a specific one. In \cref{sec:intro,sec:overview},
``the agent'' is the deployed system, which the LLM agent models. The reference agent is the \plan{} semantics (\cref{sec:plan}), the
LLM agent adds injected calls (\cref{sec:inject}), and either runs under
a monitor (\cref{sec:compose}).

\subsection{The plan language}\label{sec:plan}

\myparagraph{Preliminaries.} The language semantics has three
parameters, the sets $\mathit{Var}$ of variables, $\mathit{Val}$ of
values and $\mathit{Tool}$ of tools, which are instantiated per example,
e.g., $\mathit{Tool} = \{\Fetch, \Inbox\}$ in \exSum{}.  A tool is the
name of an external service, acting as a \emph{source} when untrusted
content is read from it and as a \emph{sink} when a request is made
on it, and may be both. 

A \emph{memory} is a function
$\mem{} \in \mathit{Mem} = \mathit{Var} \to \mathit{Val}$, and
$\mem{}[x \mapsto v]$ updates $\mem{}$ at $x$. For a set $A$, $A^{*}$ is the set of finite lists over $A$, with the
empty list $[]$. For lists $L, L' \in A^{*}$ and $a \in A$, $|L|$ is the
length of $L$, $L[i]$ is its $i$-th element, counted from $0$,
$L \cdot a$ appends $a$ to $L$, and $L \mathbin{+\!\!+} L'$ concatenates $L$ and $L'$. $A^{\omega}$ is the set of infinite
streams over $A$, and $a \cdot S$ is the stream with head $a$ and tail
$S \in A^{\omega}$. Finally,
$A_{\bot} = A \cup \{\bot\}$ adds an element $\bot \notin A$ for no value, and
$\mathit{Bool} = \{\mathit{true}, \mathit{false}\}$.

\myparagraph{Commands.} With $x \in \mathit{Var}$, $t \in \mathit{Tool}$ and
$n_1, n_2 \in \mathbb{N}$, the commands are
\begin{align*}
c & ::=  \ReadT\ x \mid \ReadU\ x\ t \mid \Call\ x\ t\ g \mid \mathsf{Assign}\ x\ f  \mid  \mathsf{IfJump}\ b\ n_1\ n_2 \mid \NewSession
\end{align*}
where $n_1$ and $n_2$ are indices into the \plan{} (\cref{def:plan}).
$\ReadT\ x$ reads the next trusted input (the user's data) into $x$.
$\ReadU\ x\ t$ reads the next untrusted item, from source tool $t$,
into $x$. $\Call\ x\ t\ g$ requests tool $t$ with arguments $g\,\mem{}$
and stores the result in $x$, where the \emph{argument function}
$g : \mathit{Mem} \to \mathit{Val}^{*}$ computes the list of
arguments, such as an email's recipient and body.
$\mathsf{Assign}\ x\ f$ stores in $x$ the value $f\,\mem{}$ of a
\emph{value function} $f : \mathit{Mem} \to \mathit{Val}$, such as a
summary of an email. $\mathsf{IfJump}\ b\ n_1\ n_2$ continues at $n_1$
or $n_2$ according to a \emph{condition}
$b : \mathit{Mem} \to \mathit{Bool}$, such as whether an email is a
meeting request (\exMeeting{}). An unconditional jump is the instance
$\mathsf{Jump}\ n \defeq \mathsf{IfJump}\ (\lambda \mem{}.\ \mathit{true})\ n\ n$
for $n \in \mathbb{N}$. No command shows a value to the user, since the
user's view is not modelled. A \plan{} that replies to the user computes
the reply with $\mathsf{Assign}$.

$\NewSession$ starts a fresh \model{} \emph{session}. A session is the
part of a run from its start or a $\NewSession$ command up to the next
$\NewSession$ or the end of the run, and the \model{}'s context holds the untrusted content read in
the current session. $\NewSession$ empties the context, so content
read in earlier sessions can no longer be obeyed, although it stays in
the memory, where the \plan{} may still use it. The monitor sees each
$\NewSession$ and may re-grant tools then. A loop
can, for instance, start a fresh session for each email it processes
(\exSession{}).

Since $f$, $g$ and $b$ are arbitrary functions rather
than syntax, expressions are shallowly embedded. The \model{}'s
intended work is expressed the same way. When the \plan{} asks the
\model{} to summarise an email or to decide whether it is a meeting
request, the answer is a function of the memory, so the \plan{} contains
it as an $f$ or a $b$. The injection model only has to describe how the \model{} deviates from the \plan{},
by making calls, or running programs, that the \plan{} does not contain (\cref{sec:threat}).

\begin{definition}[Plan]\label[definition]{def:plan}
  A \emph{plan} is a finite list of commands
  $\plansf = [c_0, \ldots, c_{N-1}]$, where $N = |\plansf|$. Commands
  are addressed by their index, so the program counter $\pc$ selects
  $\plansf[\pc] = c_{\pc}$, jump targets are indices, and the \plan{}
  has ended when $\pc \geq N$.
\end{definition}
Both agents run the same \plan{} and differ only in whether the
\model{} may deviate.

\myparagraph{Model.} 
\label{def:model}
The \model{} is the LLM inside the agent, whose states cannot be
accessed by the \plan{}. The \model{}'s behaviour is not specified,
so \plant{} describes it only by the choices that
matter for injection, which are left as parameters and chosen by
\attacker{}. Let $\Omega$ denote the set of the \model{}'s states. The \model{}'s \emph{context} is the list of untrusted items read in the current session, which it may treat as instructions. Formally, a
\emph{model} is a tuple
$(\mathit{parse}, \mathit{pick}, \mathit{onext}, o_0)$, where
$o_0 \in \Omega$ denotes its initial state, and $\mathit{onext} : \Omega \to \Omega$
advances the state each time the agent executes a
command of the \plan{} or makes an injected call (see \cref{fig:sos}).\looseness=-1

The parser
$\mathit{parse}: \mathit{Val} \to (\mathit{Tool} \times (\mathit{Mem}
\to \mathit{Val}^{*}))_{\bot}$ takes an untrusted item $v$, such
as the text of an email, and decides whether the \model{} would treat
it as an instruction, and which call that instruction asks for. If
$\mathit{parse}\ v = (t, g)$, the item $v$ contains
an \emph{instruction} to call tool $t$ with the arguments $g\,\mem{}$,
computed from the memory $\mem{}$ when the call is made.

The function $\mathit{pick}: \Omega \to \mathbb{N}_{\bot}$
decides, in a given state, whether the \model{} now obeys an item of
its context. The context may hold several untrusted items, so
$\mathit{pick}$ also chooses which one, by its index.
If $\mathit{pick}\ o = i$, the \model{} in state $o$
obeys item $i$ of its context, and if $\mathit{pick}\ o = \bot$
it obeys nothing. The pair
$(\mathit{pick}, \mathit{onext})$ is the \emph{oracle}.
The reference agent uses only $\mathit{onext}$ and $o_0$. The parser
and the oracle determine the injected calls of the LLM agent
(\cref{sec:inject}). Since $\mathit{pick}$ sees only the state, which
depends only on the number of steps taken, the oracle is a schedule
fixed in advance, whereas the arguments $g\,\mem{}$ may depend on the whole
memory, including the secrets.

\myparagraph{Configurations.} An agent configuration is a tuple
$\kappa = \langle \pc, \mem{}, T, U, C, o, \pi \rangle$, comprising the
program counter $\pc$; the memory $\mem{} \in \mathit{Mem}$; two input
streams $T, U \in \mathit{Val}^{\omega}$; the \model{}'s context
$C \in \mathit{Val}^{*}$; the \model{}'s state $o \in \Omega$; and
a pending request with $x \in \mathit{Var}$ and $v \in \mathit{Val}$,
\[
  \pi \;::=\; \emptyset \;\mid\; (x, \bot) \;\mid\; (x, v),
\]
which is empty, a call awaiting its result to be stored in $x$, or a
read of content $v$ into $x$ awaiting the monitor's decision. The
monitor's response resolves it (\cref{sec:compose}).
$\ReadT$ and $\ReadU$ consume the head of $T$ and $U$, and $T$
and $U$ are infinite, so a read never blocks and a run ends only with
the \plan{}.


Initially, we assume $\pc = 0$, $C = []$, $o = o_0$ and
$\pi = \emptyset$. The memory and both streams are arbitrary.

\myparagraph{Reference agent.} The reference agent is a labelled
transition relation $\kappa \xrightarrow{\ell}_{\Van} \kappa'$ on
configurations
$\kappa = \langle \pc, \mem{}, T, U, C, o, \pi \rangle$, with labels
\[
  \ell ::= \mathsf{AReadT}\ v \mid \mathsf{AReadU}\ t\ v \mid \mathsf{ACall}\ t\ \mathit{args} \mid \tau \mid \mathsf{ANew}
\]
as defined in \cref{fig:sos}. The labels are the agent's interface to
the monitor, not what \attacker{} observes.

Each step executes $\plansf[\pc]$ and advances the \model{}'s state to
$o' = \mathit{onext}\ o$. \textsc{ReadU} emits the request
$\mathsf{AReadU}\ t\ v$ but does not store $v$. It records the pending
read $(x, v)$, and $v$ enters the memory and the context only if the
monitor allows the read. \textsc{Call} records the pending call $(x, \bot)$, whose result is stored in $x$
later. \textsc{Assign} and \textsc{IfJump} are internal steps $\tau$,
and \textsc{New} empties the context. The reference never consults $o$,
and advances it only because the LLM agent reuses these rules, so runs
of equal length have equal \model{} states. A configuration with
$\pc \geq |\plansf|$ is final.

\begin{figure}[t]
\footnotesize
\begin{mathpar}[\lineskip=0.5em plus 0.2em]
\inferrule[ReadT]
  {\plansf[\pc] = \ReadT\ x \\ T = v \cdot T'}
  {\langle \pc, \mem{}, T, U, C, o, \pi \rangle \xrightarrow{\mathsf{AReadT}\ v}_{\Van} \\\\ \langle \pc{+}1, \mem{}[x \mapsto v], T', U, C, o', \pi \rangle}
\and
\inferrule[ReadU]
  {\plansf[\pc] = \ReadU\ x\ t \\ U = v \cdot U' }
  {\langle \pc, \mem{}, T, U, C, o, \emptyset \rangle \xrightarrow{\mathsf{AReadU}\ t\ v}_{\Van}\\\\ \langle \pc{+}1, \mem{}, T, U', C, o', (x, v) \rangle}
\and
\inferrule[Call]
  {\plansf[\pc] = \Call\ x\ t\ g }
  {\langle \pc, \mem{}, T, U, C, o, \emptyset \rangle \xrightarrow{\mathsf{ACall}\ t\ (g\,\mem{})}_{\Van} \\\\ \langle \pc{+}1, \mem{}, T, U, C, o', (x, \bot) \rangle}
\and
\inferrule[Assign]
  {\plansf[\pc] = \mathsf{Assign}\ x\ f}
  {\langle \pc, \mem{}, T, U, C, o, \pi \rangle \xrightarrow{\tau}_{\Van} \\\\ \langle \pc{+}1, \mem{}[x \mapsto f\,\mem{}], T, U, C, o', \pi \rangle}
\and
\inferrule[IfJump]
  {\plansf[\pc] = \mathsf{IfJump}\ b\ n_1\ n_2 \\\\ n = (\text{if } b\,\mem{} \text{ then } n_1 \text{ else } n_2)}
  {\langle \pc, \mem{}, T, U, C, o, \pi \rangle \xrightarrow{\tau}_{\Van} \langle n, \mem{}, T, U, C, o', \pi \rangle}
\and
\inferrule[New]
  {\plansf[\pc] = \NewSession}
  {\langle \pc, \mem{}, T, U, C, o, \pi \rangle \xrightarrow{\mathsf{ANew}}_{\Van} \langle \pc{+}1, \mem{}, T, U, [], o', \pi \rangle}
\end{mathpar}
\hrule height 0.8pt
\begin{mathpar}[\lineskip=0.5em plus 0.2em]
\inferrule[Plan]
  {\mathit{injCall}\ \kappa = \bot \\ \kappa \xrightarrow{\ell}_{\Van} \kappa'}
  {\kappa \xrightarrow{\ell}_{\Opt} \kappa'}
  \and
  \inferrule[Inject]
  {\pc < |\plansf|  \\ \mathit{injCall}\ \langle \pc, \mem{}, T, U, C, o, \emptyset \rangle = (t, g)}
  {\langle \pc, \mem{}, T, U, C, o, \emptyset \rangle \xrightarrow{\mathsf{ACall}\ t\ (g\,\mem{})}_{\Opt} \langle \pc, \mem{}, T, U, C, o', \emptyset \rangle}
\end{mathpar}
\caption{Above the line, the reference agent $\xrightarrow{}_{\Van}$,
  which executes the \plan{}. Below, the LLM agent
  $\xrightarrow{}_{\Opt}$ (\cref{def:llm}), which either takes a step
  of the \plan{} or makes the injected call (\cref{def:injcall})
  leaving the program counter unchanged. In every rule, $o'$
  abbreviates $\mathit{onext}\ o$ (\cref{def:model}), so each step
  advances the \model{}'s state.}\label{fig:sos}
\end{figure}

\begin{example}[\exSum{} continued]\label[example]{ex:sum-plan}
  The \plan{} is
  \[
    [\,\ReadT\ \vsecret,\ \ReadU\ \vmail\ \Inbox,\ \mathsf{Assign}\ \vsummary\ (\lambda \mem{}.\ \mem{}\,\vmail)\,],
  \]
  with variables $\vsecret$, $\vmail$ and $\vsummary$, and tools
  $\Fetch$ and $\Inbox$. 
  The
  \model{}'s state is a step counter, so $\Omega = \mathbb{N}$,
  $o_0 = 0$ and $\mathit{onext}\ o \defeq o + 1$. The parser and the
  oracle's $\mathit{pick}$ function are
  \begin{align*}
    \mathit{parse}\ v & \defeq  \text{if } v = \vemail \text{ then } (\Fetch, \lambda \mem{}.\ [\mem{}\,\vsecret]) \text{ else } \bot
    \\
    \mathit{pick}\ o & \defeq \text{if } o = 2 \text{ then } 0 \text{ else } \bot
  \end{align*}
  Let $T \defeq s \cdot T'$ and $U \defeq \vemail \cdot U'$, so the
  secret is $s$. The run then proceeds as follows.
  \begin{description}[leftmargin=1em,itemsep=0pt,parsep=0pt,topsep=2pt]
  \item [When $o = 0$:] $\ReadT$ reads the secret into $\vsecret$.
  \item [When $o = 1$:] $\ReadU$ requests the email. If allowed, the email
    is stored in $\vmail$ and the context becomes $[\vemail]$.
  \item [When $o = 2$:] $\mathit{pick}$ chooses the item at index $0$ of
    the context, the $\vemail$ read when $o = 1$, and
    $\mathit{parse}$ reads it as an instruction to call \Fetch{} with
    the secret held in $\vsecret$. The reference ignores this and
    executes
    $\mathsf{Assign}\ \vsummary\ (\lambda \mem{}.\ \mem{}\,\vmail)$,
    while the LLM agent, whose configuration $\kappa$ now has
    $\mathit{injCall}\ \kappa = (\Fetch, \lambda \mem{}.\ [\mem{}\,\vsecret])$
    (\cref{def:injcall}), requests
    $\Fetch\ [s]$ instead.
  \end{description}
\end{example}

\subsection{Injection models}\label{sec:inject}

This section describes \emph{one-call} injection; \cref{sec:threat}
outlines the other two models.

\begin{definition}[Injected call]\label[definition]{def:injcall}
The injected call of a configuration $\kappa = \langle \pc, \mem{}, T, U, C, o, \pi \rangle$ is
\[
  \mathit{injCall}\ \kappa \defeq
  \text{if } \mathit{pick}\ o = i \text{ and } i < |C| \text{ then } \mathit{parse}\ (C[i]) \text{ else } \bot
\]
\end{definition}

\begin{definition}[LLM agent]\label[definition]{def:llm}
The \emph{LLM agent} is the labelled transition relation $\xrightarrow{}_{\Opt}$ on configurations given by the rules \textsc{Inject} and \textsc{Plan} of \cref{fig:sos}. Its final configurations are the reference's.
\end{definition}
\noindent
In a non-final configuration, the LLM agent makes the injected call if
there is one, and otherwise takes the \plan{}'s step. An injected step
leaves the program counter unchanged, so the \plan{}'s step is only
postponed. Moreover, it leaves the pending request empty, so the
call's result is not
stored. 

The reference is the LLM agent with a parser that recognises nothing,
so the only thing to justify about the reference is the \plan{}
itself.\looseness=-1

\begin{lemma}[Reference agent]\label[lemma]{lem:ref}
  If $\mathit{parse}\ v = \bot$ for all $v$, then $\;{\xrightarrow{}_{\Opt}}=  {\xrightarrow{}_{\Van}}$.
\end{lemma}

\subsection{Composition with a monitor}\label{sec:compose}

In this section, we define the runtime monitor, which allows or denies
every request and untrusted read of an agent. We describe
the agent-monitor composition, including how the agent receives the
monitor's response, and end with the capability monitor and the
running example.

\begin{definition}[Monitor]\label[definition]{def:monitor}
  Let $\Sigma$ denote the set of monitor states, let
  $\mathit{Resp} = \{\mathsf{Allowed}, \mathsf{Denied}\}$, and let
  $\mathit{Lab}$ be the set of agent labels $\ell$ (\cref{sec:plan}). A
  \emph{monitor} over $\Sigma$ is a triple
  $(\mu_0, \mathit{mdecide}, \mathit{mnext})$ with an initial state
  $\mu_0 \in \Sigma$, a decision function
  $\mathit{mdecide} : \Sigma \to \mathit{Tool} \to \mathit{Val}^{*}
  \to \mathit{Resp} \times \Sigma$, and a state update function
  $\mathit{mnext} : \Sigma \to \mathit{Lab} \to \Sigma$.
\end{definition}
The decision $\mathit{mdecide}\ \mu\ t\ \mathit{args} = (r, \mu')$
answers a request on tool $t$ with arguments $\mathit{args}$ by a
response $r$ and a new state $\mu'$. The update
$\mathit{mnext}\ \mu\ \ell$ returns the monitor state after an agent
step with label $\ell$, and may, for instance, revoke a tool after an untrusted read. The
composite applies $\mathit{mnext}$ to trusted inputs, allowed untrusted
reads and fresh sessions (\cref{fig:compose}), and a call changes the
monitor state only through $\mathit{mdecide}$.
The agent receives the monitor's response
$r \in \mathit{Resp}$ to the pending request of
$\kappa = \langle \pc, \mem{}, T, U, C, o, \pi \rangle$ as follows,
where $\mathit{ok}, \mathit{fail} \in \mathit{Val}$ are fixed values:
\[
\mathit{recv}\ r\ \kappa \defeq
\begin{cases}
\langle \pc, \mem{}[x \mapsto v], T, U, C \cdot v, o, \emptyset \rangle & \text{if } \pi = (x, v) \wedge r = \mathsf{Allowed},\\
\langle \pc, \mem{}[x \mapsto \mathit{ok}], T, U, C, o, \emptyset \rangle & \text{if } \pi = (x, \bot) \wedge r = \mathsf{Allowed},\\
\langle \pc, \mem{}[x \mapsto \mathit{fail}], T, U, C, o, \emptyset \rangle & \text{if } \pi \in \{(x, v), (x, \bot)\} \wedge r = \mathsf{Denied},\\
\kappa & \text{if } \pi = \emptyset.
\end{cases}
\]
The result that a call stores in its variable is thus $\mathit{ok}$ or
$\mathit{fail}$, which records only whether the monitor allowed the
call. In particular, no data comes back from the tool, so
\attacker{}'s content reaches the agent only through untrusted
reads. 

A \emph{composite configuration} is a pair $(\kappa, \mu)$ of an agent
configuration and a monitor state. A composite step takes one agent
step. If the step is a request, the monitor decides it and the agent
receives the response by $\mathit{recv}$, all in the same step. The
rules are given in \cref{fig:compose}, where $X \in \{\Opt, \Van\}$.
Since an agent step has one label, a composite step emits at most one
\emph{effect} $e$, which is a trusted input $v$, an allowed read $(t, v)$, an
allowed call $(t, \mathit{args})$, or nothing ($\epsilon$).

For an untrusted read (\textsc{C-ReadU} and \textsc{C-ReadUDenied}),
the monitor evaluates $\mathit{mdecide}\ \mu\ t\ []$, with the source tool $t$ and no arguments. So its decision cannot depend on the content $v$ being read. If the read is allowed, $\mathit{recv}$ stores $v$ and adds it to the context, and if it is denied, $v$ is dropped. A denied
request, read or call, is not carried out, and its step has effect
$\epsilon$.

Every request rule of \cref{fig:compose} applies $\mathit{recv}$ to the
agent's new configuration, which clears the pending request that the
agent step has just recorded. Since initial configurations have
$\pi = \emptyset$, every configuration that a composite run reaches
has $\pi = \emptyset$ too, so the rules \textsc{ReadU}, \textsc{Call}
and \textsc{Inject} of \cref{fig:sos}, which require $\pi = \emptyset$,
never block a composite step. The composite is final when the agent
is. The LLM agent and the reference are composed with the \emph{same}
monitor and initial monitor state.

\begin{figure}[t]
\footnotesize
\begin{mathpar}[\lineskip=0.5em plus 0.2em]
\inferrule[C-ReadT]
  {\kappa \xrightarrow{\mathsf{AReadT}\ v}_{X} \kappa'}
  {(\kappa, \mu) \xRightarrow{v} (\kappa', \mathit{mnext}\ \mu\ (\mathsf{AReadT}\ v))}
\and
\inferrule[C-ReadU]
  {\kappa \xrightarrow{\mathsf{AReadU}\ t\ v}_{X} \kappa' \\ \mathit{mdecide}\ \mu\ t\ [] = (\mathsf{Allowed}, \mu'')}
  {(\kappa, \mu) \xRightarrow{(t, v)} (\mathit{recv}\ \mathsf{Allowed}\ \kappa',  \mathit{mnext}\ \mu''\ (\mathsf{AReadU}\ t\ v))}
\and
\inferrule[C-ReadUDenied]
  {\kappa \xrightarrow{\mathsf{AReadU}\ t\ v}_{X} \kappa' \\ \mathit{mdecide}\ \mu\ t\ [] = (\mathsf{Denied}, \mu')}
  {(\kappa, \mu) \xRightarrow{\epsilon} (\mathit{recv}\ \mathsf{Denied}\ \kappa', \mu')}
\and
\inferrule[C-Call]
  {\kappa \xrightarrow{\mathsf{ACall}\ t\ \mathit{args}}_{X} \kappa' \\ \mathit{mdecide}\ \mu\ t\ \mathit{args} = (\mathsf{Allowed}, \mu')}
  {(\kappa, \mu) \xRightarrow{(t, \mathit{args})} (\mathit{recv}\ \mathsf{Allowed}\ \kappa', \mu')}
\and
\inferrule[C-CallDenied]
  {\kappa \xrightarrow{\mathsf{ACall}\ t\ \mathit{args}}_{X} \kappa' \\ \mathit{mdecide}\ \mu\ t\ \mathit{args} = (\mathsf{Denied}, \mu')}
  {(\kappa, \mu) \xRightarrow{\epsilon} (\mathit{recv}\ \mathsf{Denied}\ \kappa', \mu')}
\and
\inferrule[C-New]
  {\kappa \xrightarrow{\mathsf{ANew}}_{X} \kappa'}
  {(\kappa, \mu) \xRightarrow{\epsilon} (\kappa', \mathit{mnext}\ \mu\ \mathsf{ANew})}
\and
\inferrule[C-Tau]
  {\kappa \xrightarrow{\tau}_{X} \kappa'}
  {(\kappa, \mu) \xRightarrow{\epsilon} (\kappa', \mu)}
\end{mathpar}
\caption{The composite step of an agent $X$ (the reference or the LLM agent) and a monitor. The label over $\Rightarrow$ is the step's effect.}\label{fig:compose}
\vspace{-10pt}
\end{figure}

\begin{example}[Capability monitor]
  \label[example]{ex:comp-with-monit}
  A capability monitor is used to model examples such as \exSum{}. Its
  parameters are the initial set $\mathcal{C}_0 \subseteq \mathit{Tool}$
  of tools the agent holds, a grant
  $\mathit{grantOn} : \mathit{Val} \to 2^{\mathit{Tool}}$ for trusted
  inputs, and the fixed set $\mathit{revokeOnU} \subseteq \mathit{Tool}$
  that an untrusted read revokes. Its states are sets of tools
  $\mathcal{C} \subseteq \mathit{Tool}$ that the agent may use, where
  $\mu_0 = \mathcal{C}_0$. Moreover,
  \[
    \mathit{mdecide}\ \mathcal{C}\ t\ \mathit{args} \defeq
    \begin{cases}
      (\mathsf{Allowed}, \mathcal{C}) & \text{if } t \in \mathcal{C},\\
      (\mathsf{Denied}, \mathcal{C}) & \text{else,}
    \end{cases}
    \quad
    \mathit{mnext}\ \mathcal{C}\ \ell \defeq
    \begin{cases}
      \mathcal{C} \cup \mathit{grantOn}\ v & \text{if } \ell = \mathsf{AReadT}\ v,\\
      \mathcal{C} \setminus \mathit{revokeOnU} & \text{if } \ell = \mathsf{AReadU}\ t\ v,\\
      \mathcal{C} & \text{else.}
    \end{cases}
  \]
  So a request is allowed iff its tool is held, a trusted input $v$ grants
  the tools $\mathit{grantOn}\ v$, and an allowed untrusted read revokes
  the fixed set $\mathit{revokeOnU}$, which is the \emph{fence}.
\end{example}

\begin{example}[\textsc{Leaky and Fenced Summariser} continued]
  \exSum{} uses the capability monitor with $\mathcal{C}_0 = \{\Fetch, \Inbox\}$,
  $\mathit{grantOn}\ v = \emptyset$ and $\mathit{revokeOnU} = \emptyset$.
  Since $\Inbox \in \mathcal{C}$, \textsc{C-ReadU} allows the read of
  the email, and since $\Fetch \in \mathcal{C}$, \textsc{C-Call} allows
  the injected $\Fetch\ [s]$, with effect $(\Fetch, [s])$, which the
  reference never produces. \exFenced{} has $\mathit{revokeOnU} = \{\Fetch\}$, so its read removes $\Fetch$ from
  $\mathcal{C}$ and \textsc{C-CallDenied} denies the call (effect $\epsilon$).
\end{example}


%% file: sections/rs.tex
\section{Relative Security for Agents}\label{sec:rs}

\myparagraph{Relative security.}\label{sec:rs-def}
Relative security~\cite{dongol2024csf,derrick2025jar} compares two
transition systems, a reference $\Van$ and an optimised system
$\Opt$. 
Here $\Opt$ is the composite of the LLM agent with a monitor, and
$\Van$ the composite of the reference with the same monitor
(\cref{sec:compose}). Given a
\emph{system} $X$ with states $S_X$, initial states
$I_X \subseteq S_X$, final states $F_X \subseteq S_X$ and a transition
relation ${\to_X} \subseteq S_X \times S_X$, a \emph{run} of $X$ is a
finite or infinite sequence $\tr = s_0 s_1 \ldots$ of states with
$s_0 \in I_X$ and $s_i \to_X s_{i+1}$ for all consecutive
$s_i, s_{i+1}$. A finite run is \emph{completed} if its last state is
in $F_X$, and every infinite run is completed. We write
$\mathit{Runs}_X$ for the finite completed runs of $X$,
$\mathit{Runs}^{\infty}_X$ for all completed runs, and $\tr[0]$ for
the first state of $\tr$. An \emph{attacker model} for $X$ gives each
run its secrets $\Sec(\tr)$, attacker actions $\Act(\tr)$ and
observations $\Obs(\tr)$, all finite or infinite sequences.
Two runs $\tr_1, \tr_2$ \emph{leak} $(\sigma_1, \sigma_2)$ if
\[
\mathit{leak}(\tr_1, \tr_2, \sigma_1, \sigma_2) \defeq
\Sec(\tr_1) = \sigma_1 \wedge \Sec(\tr_2) = \sigma_2 \wedge
\Act(\tr_1) = \Act(\tr_2) \wedge \Obs(\tr_1) \neq \Obs(\tr_2).
\]
Initial composite states have the form $(\kappa, \mu_0)$ with
$\kappa = \langle 0, \mem{}, T, U, [], o_0, \emptyset\rangle$
(\cref{sec:plan}). Write $\kappa.\mem{}$, $\kappa.T$ and $\kappa.U$
for the memory and the two input streams of $\kappa$. Two initial
composite states \emph{correspond} if they have the same memory and the
same trusted inputs,
\[
\mathit{corr}((\kappa_v, \mu_0), (\kappa, \mu_0)) \defeq
\kappa_v.\mem{} = \kappa.\mem{} \wedge \kappa_v.T = \kappa.T.
\]
The untrusted inputs $\kappa_v.U$ and $\kappa.U$ may differ, since they
are \attacker{}'s actions. The program model and the dual-LLM
composites use the same condition on their configurations. Composition and stuttering simulation
(\cref{sec:compose-method,sec:stutter}) hold for any two agents over a
common state type composed with a monitor, so there, as in the AFP,
$\mathit{corr}$ is a parameter. Lockstep and block simulation are
proved for the plan language with any monitor.

\begin{definition}[Relative security]\label[definition]{def:rsecure}
We say that $\Opt$ is \emph{relatively secure} with respect to $\Van$ if
every leak of $\Opt$ is a leak of $\Van$ from corresponding initial
states,
\begin{multline*}
\forall\, \tr_1, \tr_2 \in \mathit{Runs}_{\Opt},\ \sigma_1, \sigma_2.\
\mathit{leak}(\tr_1, \tr_2, \sigma_1, \sigma_2) \Longrightarrow {}\\
\exists\, \trv_1, \trv_2 \in \mathit{Runs}_{\Van}.\
\mathit{leak}(\trv_1, \trv_2, \sigma_1, \sigma_2) \wedge
\mathit{corr}(\trv_1[0], \tr_1[0]) \wedge \mathit{corr}(\trv_2[0], \tr_2[0]).
\end{multline*}
\end{definition}

Reference runs must agree with each other on
actions, but not with the LLM-agent runs, so the reference may use a
different attacker input to produce the same leak. This makes \exMeeting{} (\cref{sec:disclosure-design}) secure.
The reference may also use a different input for each pair of secrets,
so relative security bounds which pairs of secrets \attacker{} can
tell apart with \emph{some} input, not what it learns from one input.
The difference matters. Suppose the reference reads one guess $g$ from
\attacker{} and replies with the bit $[s = g]$ for a secret $s$, and an
injection makes the LLM agent also send~$s$. For each pair
$s_1 \neq s_2$ the reference with guess $s_1$ tells them apart, so the
injected disclosure of all of~$s$ is relatively secure, although each
reference run reveals one bit. Relative security thus preserves the
task's indistinguishability over all of \attacker{}'s inputs, not the
information available in one query, session or budget. For the latter
the reference's inputs must not depend on the secrets.

\begin{definition}[Uniform relative security]\label[definition]{def:ursecure}
$\Opt$ is \emph{uniformly} relatively secure if there is a function
$\alpha$ on action sequences such that every leak
$(\tr_1, \tr_2, \sigma_1, \sigma_2)$ of $\Opt$ is a leak of reference
runs $\trv_1, \trv_2$ from corresponding initial states with
$\Act(\trv_i) = \alpha(\Act(\tr_i))$.
\end{definition}
\noindent 
Since $\alpha$ is fixed before the runs and secrets, two LLM-agent runs
with equal actions are matched by reference runs with equal actions,
whatever the secrets. 
Uniform relative security implies relative security and is strictly
stronger. In the guessing example above, $\alpha$ fixes the reference's
guess $g'$, and the reference then gives the same reply for any two
secrets other than $g'$, which the LLM agent tells apart.

Relative security on \emph{possibly infinite runs} is \cref{def:rsecure}
with $\mathit{Runs}^{\infty}$ in place of $\mathit{Runs}$. A
conversational loop never reaches a final state, so it has no finite
completed runs and is trivially secure on them. Its verdicts are
therefore given on possibly infinite runs, which need the extra
condition of \cref{sec:loops}.

\myparagraph{The attacker models.}\label{sec:attackers}
Our attacker models are \emph{statewise}, as in the AFP. Some states \emph{produce a secret}, and some are \emph{interaction points}, at which \attacker{} may act or observe, each with an \emph{interaction}, a pair of an action and an observation. $\Sec(\tr)$ is the list of secrets produced along $\tr$, and $\Act(\tr)$ and $\Obs(\tr)$ are the lists of actions and of observations at its interaction points.

Fix the set of tools visible to \attacker{}. For a composite configuration $(\kappa, \mu)$, write $\ell$ for the label of the agent's next step from $\kappa$, and $\mathit{allowed}\ \mu\ t\ \mathit{args}$ if $\mathit{mdecide}\ \mu\ t\ \mathit{args}$ allows the request.

\begin{definition}[Event-level attacker]\label[definition]{def:event}
A configuration $(\kappa, \mu)$ produces secret $v$ iff $\ell = \mathsf{AReadT}\ v$. It is an interaction point iff $\ell = \mathsf{AReadU}\ t\ v$ and $\mathit{allowed}\ \mu\ t\ []$, with interaction $(v, \bot)$, or $\ell = \mathsf{ACall}\ t\ \mathit{args}$ with $t$ visible and $\mathit{allowed}\ \mu\ t\ \mathit{args}$, with interaction $(\bot, (t, \mathit{args}))$.
\end{definition}

\attacker{}'s actions are thus the contents of the allowed untrusted reads, and its observations the allowed visible calls with their arguments. Each interaction has $\bot$ on one side, so $\Act$ and $\Obs$ keep their positions, and together they show whether a read came before or after a call. The attacker does not see steps that are not interactions. Denied calls, trusted inputs and internal steps are invisible to it.

Since $\Act$ has a $\bot$ for each allowed visible call, two runs are
compared only if they have the same \emph{interaction pattern}, that
is, the same number of calls reaches \attacker{} between consecutive
reads. The event-level attacker therefore sees the contents and order
of the messages it receives, but a difference in their \emph{number}
is not by itself an event-level leak. For example, suppose an
injection fetches a permitted URL if the secret is~0 and a forbidden
one otherwise, and the \plan{} then makes a fixed allowed call that
\attacker{} sees, a completion message. \attacker{} receives the fetch
and the completion message in one run and only the completion message
in the other, and a real recipient can tell these apart. But the two
runs have different interaction patterns, so they are never compared,
and the event level does not report the leak (\exAllow{},
\cref{sec:examples}). Event-level verdicts are thus conditional on
matching interaction patterns. Where both composites have alignments
(\cref{def:aligned}), \cref{thm:transfer} equates them with the
step-level verdicts. The step level detects the \exAllow{}
channel. More general channels through the number of messages or
termination need an observation model that compares the relevant runs,
which neither level does in general.

\begin{definition}[Step-level attacker]\label[definition]{def:step}
Secrets are as in \cref{def:event}. Every non-final state is an interaction point. Its action is $v$ if it is an allowed untrusted read of $v$, and $\bot$ otherwise. Its observation is $(t, \mathit{args})$ if its label is $\mathsf{ACall}\ t\ \mathit{args}$ with $t$ visible and the request allowed, and $\bot$ otherwise.
\end{definition}

The step-level attacker additionally sees how many steps pass between interactions, but it compares only runs that take the same number of steps and read at the same steps. It is therefore not simply a stronger observer than the event-level one, since it also restricts which runs are compared. For instance, if the secret decides whether the agent takes an extra internal step and an injection sends one bit of the secret, the agent is relatively secure at the step level, which never compares runs with different bits, but not at the event level, which does, and relative security at one level need not carry over to the other, and \cref{thm:transfer} gives a condition under which it does. Lockstep, block simulation and disproof (\cref{sec:lockstep,sec:block,sec:disproof}) work at this level. A third attacker model, used in \cref{sec:examples}, sees more than allowed requests.

\begin{definition}[Attackers with views]\label[definition]{def:views}
Fix two observation functions, a \emph{denial view} $\mathit{dview}$ on tools and arguments, and a \emph{read view} $\mathit{rview}$ on tools. Either may return $\bot$. The attacker is the step-level one, except that a state whose label is $\mathsf{ACall}\ t\ \mathit{args}$ and whose request is denied has observation $\mathit{dview}(t, \mathit{args})$, and an allowed untrusted read from $t$ has observation $\mathit{rview}(t)$.
\end{definition}

The denial view models a monitor whose refusals are visible, as an error message, a rate-limit counter or a public log. The read view models sources \attacker{} runs, such as a page whose server sees which requests it serves. Since compared runs read at the same steps, the read view adds only which source each read uses. With both views always $\bot$ the attacker is the step-level one.

\myparagraph{Transfer between the levels.}\label{sec:transfer}
A run's event-level actions and observations are its step-level ones with the empty interactions removed. The levels therefore coincide when the \emph{pattern} of interactions depends only on the attacker's actions.\looseness=-1

\begin{definition}[Alignment]\label[definition]{def:aligned}
A relation $R$ on composite states of one system is an \emph{alignment} if (i) it holds for all pairs of initial states, (ii) related states are final together and are event-level interaction points together, and (iii) steps with equal step-level actions preserve the relation.
\end{definition}

\begin{theorem}[Transfer]\label{thm:transfer}
If both composites have alignments, then relative security at the step level holds iff it holds at the event level, and likewise on possibly infinite runs.
\end{theorem}


So, where both composites have alignments, a step-level verdict by unwinding, secure or insecure, becomes an event-level verdict. Without alignment the levels can differ. \exCallApp{} is secure at the event level and leaks through timing at the step level (\cref{sec:approving-tool-or,sec:examples}).

\myparagraph{Disclosure bounds.}\label{sec:disclosure}
Relative security bounds what an injection adds, not what the \plan{}
discloses. A disclosure bound states how much the \plan{} may
disclose.

\begin{definition}[Disclosure bound]\label[definition]{def:discloses}
$\Van$ \emph{discloses at most} $\delta$ if any two runs $\trv_1, \trv_2 \in \mathit{Runs}_{\Van}$ with $\Act(\trv_1) = \Act(\trv_2)$ and $\delta(\Sec(\trv_1)) = \delta(\Sec(\trv_2))$ have $\Obs(\trv_1) = \Obs(\trv_2)$. $\Opt$ discloses at most $\delta$ when the same holds for $\mathit{Runs}_{\Opt}$.\looseness=-1
\end{definition}

\begin{theorem}[Disclosure bounds carry over]\label{thm:discloses}
If $\Opt$ is relatively secure and $\Van$ discloses at most $\delta$, then $\Opt$ discloses at most $\delta$, and likewise on possibly infinite runs, with $\mathit{Runs}^{\infty}$ for $\mathit{Runs}$.
\end{theorem}

By relative security, a leak of $\Opt$ between secrets $\sigma_1, \sigma_2$ with $\delta(\sigma_1) = \delta(\sigma_2)$ would be a leak of $\Van$ between the same secrets, which the bound excludes. The bound is checked on the reference, where injection plays no role, and carried to the LLM agent for every \model{} behaviour under which relative security holds. Since the runs may start from any memories, the bound treats the initial memory as secret. For a \plan{} that makes no calls, $\delta$ can be constant, so every relatively secure agent with such a \plan{}, \exFenced{} for instance, is noninterferent at the event level.


%% file: sections/methods.tex
\section{Fence Theorems and Proof Methods}\label{sec:methods}\label{sec:defence}

Each verdict in \cref{sec:examples} is proved by a generic theorem
applied to the example. Most
security verdicts are proved at the step level, by the existing unwinding
theorem~\cite{derrick2025jar,afp-relsec} where it applies, and transferred to the event level
(\cref{thm:transfer}). The fence theorems and stuttering give
event-level verdicts directly.

\subsection{The fence theorems}\label{sec:fence}\label{sec:loops}

Injected calls are requests the reference never makes. If the monitor
silently denies every call an injection can make, from the first
untrusted read on, the injections are harmless, whatever the \plan{}.
A \emph{fence} is the capability monitor of \cref{ex:comp-with-monit}
with every tool an injection can call in $\mathit{revokeOnU}$, so it
revokes these tools at the first untrusted read.

\begin{theorem}[Revoke on untrusted read]\label{thm:revoke}
If no trusted input grants a tool in $\mathit{revokeOnU}$, the plan language composed with a fence is relatively secure at the event level, for every \plan{}, every parser whose instructions call only tools in $\mathit{revokeOnU}$, and every oracle.
\end{theorem}

The \plan{} is unchanged, and its own requests are decided identically
in both systems. In \exFenced{}, with $\mathit{revokeOnU} = \{\Fetch\}$,
the read of the email in \cref{fig:e1e2} revokes $\Fetch$ and makes the
context non-empty, so the injected $\Fetch\ [s]$ is denied silently
while the reference waits at $\mathsf{Assign}$. The theorem also covers
\plan{}s that disclose by design, for instance one that sends the
calendar through \Send{} while injections can call only \Fetch{}, and
such agents are not noninterferent. The theorem is for one-call
injection, and for injected programs a fence suffices only in a
sandbox (\cref{thm:sandbox}). Both are proved by the composition method of \cref{sec:compose-method}, with no proof per \plan{}.

The theorem needs the fenced tools never to be granted back, but a
loop that serves one request after another must grant them again. Doing
so in the same context leaks, since the user's next report grants
$\Send$ again while the context still holds the previous email
(\exStale{}). The session fence (\cref{app:monitors}) re-grants the
tools only at a fresh \model{} session. A loop never ends, so its
verdict is relative security on possibly infinite runs. This needs the oracle to be \emph{fair}, that is, the
\model{} cannot obey for ever. Formally, a measure on agent states
decreases at every injected call.

\begin{theorem}[Fence on possibly infinite runs]\label{thm:revoke-inf}
Under the hypotheses of \cref{thm:revoke}, the plan language composed with a fence is relatively secure at the event level on possibly infinite runs, for every \plan{}, every such parser and every fair oracle. The same holds for the session fence.
\end{theorem}

Fairness cannot be dropped (\cref{sec:rs}),
although the stall adds nothing to what \attacker{} sees beyond the
\plan{}'s own send of the first secret. Relative security fails there only
because it asks the reference to reproduce the secret sequence exactly,
and a stalled run reads fewer secrets than any completed reference run.
At the step level, where timing is observed, the existing unwinding theorem gives
step-level versions of \cref{thm:revoke,thm:revoke-inf} if, in addition, the \plan{}'s control does not depend on
the secrets, so that its timing and when untrusted content enters the
context depend only on \attacker{}'s input. \exSession{} has a fair
oracle and meets this condition.

\subsection{Composition}\label{sec:compose-method}

The fence theorems rest on one simulation. The LLM agent differs from the reference only by its injected requests.
If the monitor denies each of them silently, an injected step changes
nothing that \attacker{} sees, and the reference can simply wait while
it happens. The method splits this argument in two. On the agent side,
the LLM agent follows the reference except for injected requests in a
class $\Dcls$, which it makes only when \emph{tainted}, after untrusted
content has entered its context. On the monitor side, the monitor
denies every request in $\Dcls$ once it is in \emph{deny mode}, which it
enters at the untrusted read. One conjunct links the two sides, so each
is proved without the other, and one monitor proof serves every \plan{}.
The method is stated for an arbitrary pair of agents with a common
state type, not only for the plan language.

Formally, the method has three parameters, a predicate $\taint$ on agent states, which for the language is ``the context is non-empty'', a class $\Dcls$ of requests the monitor promises to deny, and a predicate $\deny$ on monitor states, \emph{deny mode}. Write $I_X$ and $F_X$ for the initial and final states of agent $X$, $a \xrightarrow{\ell}_{\Opt} a'$ for a step of the LLM agent, $a \xrightarrow{\ell}_{\Van} a'$ for one of the reference, $\xrightarrow{\tau}^{*}$ for a sequence of internal reference steps, and $\mathit{recv}\ r\ a$ for delivering the monitor's response $r$ to the agent. For a step labelled $\ell$ to $a'$, let
$\mathit{post}\ \ell\ a'\ r = \mathit{recv}\ r\ a'$ if $\ell$ is a
request or an untrusted read, and $\mathit{post}\ \ell\ a'\ r = a'$
otherwise. For a relation $\mathit{RA}$ from LLM-agent states to reference states, let $\mathit{after}(\ell, a', a_v')$ hold if
$\mathit{RA}\ (\mathit{post}\ \ell\ a'\ r)\ (\mathit{post}\ \ell\ a_v'\ r)$
for every $r \in \mathit{Resp}$.

\begin{definition}[Simulation up to $\Dcls$]\label[definition]{def:agent-ok}
A relation $\mathit{RA}$ from LLM-agent states to reference states is a
\emph{simulation up to $\Dcls$} if, for all LLM-agent states $a, a'$,
reference states $a_v$ and labels $\ell$,
\begin{enumerate}
\item for every $a \in I_{\Opt}$ there is $a_v \in I_{\Van}$ with
  $\mathit{corr}(a_v, a) \wedge \mathit{RA}\ a\ a_v$;
\item if $\mathit{RA}\ a\ a_v$ and $a \xrightarrow{\ell}_{\Opt} a'$,
  then one of the following holds:
  \begin{enumerate}[label=(\alph*)]
  \item $\ell = \tau \wedge \mathit{RA}\ a'\ a_v$;
  \item $\ell = \mathsf{ACall}\ t\ \mathit{args} \wedge \Dcls\ t\ \mathit{args}
    \wedge \taint\ a \wedge
    \mathit{RA}\ (\mathit{recv}\ \mathsf{Denied}\ a')\ a_v$ for some
    $t, \mathit{args}$;
  \item $\ell = \mathsf{AReadU}\ t\ v \wedge \Dcls\ t\ [] \wedge \taint\ a
    \wedge \mathit{RA}\ (\mathit{recv}\ \mathsf{Denied}\ a')\ a_v$ for
    some $t, v$;
  \item $a_v \xrightarrow{\tau}^{*} a_v^1 \xrightarrow{\ell}_{\Van} a_v'
    \wedge \mathit{after}(\ell, a', a_v')$ for some $a_v^1, a_v'$;
  \end{enumerate}
\item if $\mathit{RA}\ a\ a_v$ and $a \in F_{\Opt}$, then
  $a_v \xrightarrow{\tau}^{*} a_v'$ for some $a_v' \in F_{\Van}$.
\end{enumerate}
\end{definition}
Case~2a lets the LLM agent take internal steps alone. Cases~2b and~2c
are the injected requests. The reference does not move, and the LLM
agent continues as if the request were denied, which the monitor will
guarantee. Case~2d matches a step of the \plan{} by the same labelled
step of the reference, after some internal steps, and keeps
$\mathit{RA}$ whatever the monitor answers. Condition~3 ensures that the
reference can finish when the LLM agent does.

Simulation up to $\Dcls$ allows injected requests only in tainted
states. The composition is sound only if the monitor is in deny mode in
every such state, so the proof keeps the link
$\taint\ a \Longrightarrow \deny\ \mu$ between agent and monitor on
every composite step. On the agent side this needs taint to switch on
only where the monitor sees it. Taint must be off initially, may
switch on only at an untrusted read that the monitor allows, and must
then stay on until a fresh session empties the context.

\begin{definition}[Sound taint]\label[definition]{def:taint-ok}
  We say that $\taint$ is \emph{sound} iff for all steps
  $a \xrightarrow{\ell}_{\Opt} a'$ and responses $r$,
\begin{enumerate*}[label=\bfseries(\arabic*)]
\item $\neg\,\taint\ a$ for initial $a$;
\item $\taint\ a \wedge \ell \neq \mathsf{ANew} \Longrightarrow
  \taint\ (\mathit{post}\ \ell\ a'\ r)$;
\item $\ell = \mathsf{ANew} \Longrightarrow \neg\,\taint\ a'$;
\item $\neg\,\taint\ a \wedge \taint\ (\mathit{post}\ \ell\ a'\ r)
  \Longrightarrow{}$ $\ell = \mathsf{AReadU}\ t\ v \wedge r = \mathsf{Allowed}$
  for some $t, v$.
\end{enumerate*}
\end{definition}

On the monitor side, the same link needs deny mode to switch on no later
than taint and to last at least as long. The monitor must therefore
enter deny mode at every untrusted read and leave it only at a fresh
session. In deny mode it must also deny each request in $\Dcls$
without changing its state. The reference never makes such a request,
so the two composites then keep the same monitor state, as the effect
simulation below requires.

\begin{definition}[Monitor obligations]\label[definition]{def:monitor-ok}
The predicate $\deny$ and the class $\Dcls$ meet the \emph{monitor
obligations} if, for all monitor states $\mu$, tools $t$, arguments
$\mathit{args}$, items $v$ and labels $\ell$,
\begin{enumerate*}[label=\bfseries(\arabic*)]
\item $\deny\ \mu \wedge \Dcls\ t\ \mathit{args} \Longrightarrow
  \mathit{mdecide}\ \mu\ t\ \mathit{args} = (\mathsf{Denied}, \mu)$;
\item $\deny\ (\mathit{mnext}\ \mu\ (\mathsf{AReadU}\ t\ v))$;
\item $\deny\ \mu \wedge \ell \neq \mathsf{ANew} \Longrightarrow
  \deny\ (\mathit{mnext}\ \mu\ \ell)$;
\item $\deny\ \mu \wedge \mathit{mdecide}\ \mu\ t\ \mathit{args} =
  (r, \mu') \Longrightarrow \deny\ \mu'$.
\end{enumerate*}
\end{definition}


Together the three conditions give the argument sketched at the start
of this section: each injected request is denied silently while the
reference waits, and every other step is matched by the same reference
step under the same monitor state. The match ignores timing, so the
theorem holds at the event level.

\begin{theorem}[Composition]\label{thm:compose}
If $\mathit{RA}$ is a simulation up to $\Dcls$ (\cref{def:agent-ok}), $\taint$ is sound (\cref{def:taint-ok}), and $\deny$ and $\Dcls$ meet the monitor obligations (\cref{def:monitor-ok}), then the LLM-agent composite is relatively secure at the event level for every set of visible tools.
\end{theorem}

The proof goes through an \emph{effect simulation}. The relation
\[
  \mathit{CR}\ (a, \mu)\ (a_v, \mu_v) \;\defeq\; \mathit{RA}\ a\ a_v \wedge \mu = \mu_v \wedge (\taint\ a \Longrightarrow \deny\ \mu)
\]
matches each LLM-agent step by a reference path with the same effects, and a final state by a reference path with no effects. Since the secrets, actions and observations of \cref{def:event} are projections of the effects other than $\epsilon$ (\cref{sec:compose}), each LLM-agent run is matched by one reference run with the same secrets, actions and observations, which is more than relative security needs.


For possibly infinite runs, \cref{def:agent-ok} holds with a measure that
decreases at every LLM-agent step the reference does not take. The
effect simulation behind \cref{thm:compose} is then fair, so the reference, which
waits during injected calls, is never left behind for good (\cref{thm:revoke-inf}).

The fence instantiates the method with $\Dcls\ t\ \mathit{args} \defeq
t \in \mathit{revokeOnU}$, with $\deny\ \mathcal{C} \defeq \mathcal{C}
\cap \mathit{revokeOnU} = \emptyset$ for the capabilities
$\mathcal{C}$, and with $\taint$ ``the context is non-empty''. For
every oracle and every parser that yields only requests in $\Dcls$, the
plan language meets \cref{def:agent-ok,def:taint-ok}, and the
capability monitor meets \cref{def:monitor-ok} if no trusted input
grants a tool in $\mathit{revokeOnU}$.

\subsection{Unwinding and lockstep}\label{sec:lockstep}

The existing unwinding theorem~\cite{derrick2025jar,afp-relsec} proves relative security from a relation $\Delta$ on four states: two runs of $\Opt$ and two of $\Van$. Informally, $\Delta$ must hold initially and, whenever it holds, the four runs end together, the $\Opt$-runs are interaction points together while their observations so far agree, and either some runs can move ahead alone (the \emph{proactive} case, bounded by a measure) or every joint step of the $\Opt$-runs with equal actions is matched by the $\Van$-runs, with different observations on the $\Van$-side whenever those on the $\Opt$-side differ, and $\Delta$ re-established. For agents, most of $\Delta$ is determined by the language, and we package it as follows.

For a composite state $s_i = (\kappa_i, \mu_i)$ with
$\kappa_i = \langle \pc_i, \mem{}_i, T_i, U_i, C_i, o_i, \pi_i \rangle$,
write $\mathit{act}(s_i)$ for its step-level action (\cref{def:step})
and $s_i \Rightarrow s_i'$ for a composite step of the LLM agent. If
$\mathit{injCall}\ \kappa_i = (t, g)$, let
$(r_i, \mu_i') = \mathit{mdecide}\ \mu_i\ t\ (g\,\mem{}_i)$ be the
monitor's decision on the injected call, and let its observation
$\mathit{io}_i$ be $(t, g\,\mem{}_i)$ if $r_i = \mathsf{Allowed}$ and
$t$ is visible, and $\bot$ otherwise.

\begin{definition}[Lockstep]\label[definition]{def:lockstep}
A relation $Q \subseteq S_{\Opt} \times S_{\Opt}$ is a \emph{lockstep
relation} if, for all $s_1, s_2$,
\begin{enumerate}
\item $s_1, s_2 \in I_{\Opt} \Longrightarrow Q\ s_1\ s_2$;
\item $Q\ s_1\ s_2 \Longrightarrow o_1 = o_2 \wedge C_1 = C_2 \wedge \pi_1 = \pi_2 = \emptyset \wedge (s_1 \in F_{\Opt} \Leftrightarrow s_2 \in F_{\Opt})$;
\item $Q\ s_1\ s_2 \wedge s_1 \notin F_{\Opt} \wedge \mathit{injCall}\ \kappa_1 \neq \bot \Longrightarrow \mu_1' = \mu_1 \wedge \mu_2' = \mu_2 \wedge \mathit{io}_1 = \mathit{io}_2$;
\item $Q\ s_1\ s_2 \wedge s_1 \Rightarrow s_1' \wedge s_2 \Rightarrow s_2' \wedge \mathit{act}(s_1) = \mathit{act}(s_2) \Longrightarrow Q\ s_1'\ s_2'$.
\end{enumerate}
\end{definition}
By condition~2, $\mathit{injCall}\ \kappa_2 = \mathit{injCall}\ \kappa_1$,
so condition~3 concerns one injected call. It
leaves the monitor unchanged and is allowed and visible
with equal arguments in both or in neither.

\begin{theorem}[Lockstep method]\label{thm:lockstep}
If the LLM-agent composite of the plan language has a lockstep relation, it is relatively secure at the step level with respect to the reference composite.
\end{theorem}

The proof instantiates the $\Delta$ of the unwinding theorem with $Q$ on the two LLM-agent runs, and keeps each reference run in its LLM-agent run's composite configuration, apart from the \model{}'s state. The reference waits during injected calls, which by condition~2 the two LLM-agent runs make at the same moments, and which by condition~3 leak nothing. For possibly infinite runs, a measure that decreases at every injected call keeps the reference from being left behind. Since injected calls need not be denied, the method also covers argument contracts that allow some of them (\exAllow{}), and monitors without the deny mode of \cref{def:monitor-ok}, such as a budget spent before the untrusted read (\exBudgetFirst{}).

\subsection{Block simulation}\label{sec:block}

In \exMeeting{} the injected reply sends the calendar right after the email is read, whereas the reference sends it only as its last step, on a meeting request. At the step level, where every non-final state is an interaction point, the existing unwinding theorem can only move the two reference runs together, and they may not end before the LLM-agent runs. So when the injected sends first tell the LLM-agent runs apart, the reference runs have sent nothing, and no relation meets the unwinding conditions. \emph{Block simulation} matches one run at a time instead.

For a composite state $s$ of either system, write $\mathit{act}(s)$
and $\mathit{obs}(s)$ for its step-level action and observation, and
$\mathit{sec}(s)$ for the list of secrets it produces. Let
$\mathcal{A}_X$ and $\mathcal{O}_X$ be the sets of step-level actions
and observations of system $X$. For a list
$\mathit{bl} = [b_0, \ldots, b_{k-1}]$ of states, $\mathit{act}(\mathit{bl})$
is the list $[\mathit{act}(b_0), \ldots, \mathit{act}(b_{k-1})]$, likewise
$\mathit{obs}(\mathit{bl})$, and $\mathit{sec}(\mathit{bl})$ joins the
lists $\mathit{sec}(b_0), \ldots, \mathit{sec}(b_{k-1})$ into one.  Write $s_v \xrightarrow{\mathit{bl}}_{\Van} s_v'$ if
the reference can go from $s_v$ to $s_v'$ through the states
$\mathit{bl}$, that is, $\mathit{bl} = [b_0, \ldots, b_{k-1}]$ with
$b_0 = s_v$ and
$b_0 \to_{\Van} \cdots \to_{\Van} b_{k-1} \to_{\Van} s_v'$, or
$\mathit{bl} = []$ and $s_v' = s_v$. A \emph{block} is such a list
$\mathit{bl}$.

\begin{theorem}[Block simulation]\label{thm:block}
Let $\Opt$ and $\Van$ be the composites of the plan language with a
monitor, let $\mathcal{Q}$ be a set of control states with $q_0 \in
\mathcal{Q}$, and let
\begin{itemize}
\item $\mathit{ra} : \mathcal{Q} \to \mathcal{A}_{\Opt} \to \mathcal{A}_{\Van}^{*}$
  give the reference's actions in the block matching an LLM-agent action;
\item $\mathit{fa} : \mathcal{Q} \to \mathcal{A}_{\Van}^{*}$ give the
  reference's actions in a final block;
\item $\mathit{nxt} : \mathcal{Q} \to \mathcal{A}_{\Opt} \to \mathcal{Q}$
  give the next control state;
\item $\mathit{ob} : \mathcal{Q} \to \mathcal{A}_{\Opt} \to
  \mathcal{O}_{\Van}^{*} \to \mathcal{O}_{\Van}^{*} \to \mathcal{O}_{\Opt}$
  give the LLM agent's observation, computed from the history of
  reference observations and the block's observations.
\end{itemize}
Let $\mathit{Sim} \subseteq \mathcal{Q} \times \mathcal{O}_{\Van}^{*}
\times S_{\Opt} \times S_{\Van}$, and suppose
\begin{enumerate}[leftmargin=*]
\item for every $s \in I_{\Opt}$ there is $s_v \in I_{\Van}$ with
  $\mathit{corr}(s_v, s)$ and $\mathit{Sim}\ q_0\ []\ s\ s_v$;
\item if $\mathit{Sim}\ q\ h\ s\ s_v$ and $s \to_{\Opt} s'$, then there
  are $\mathit{bl}$ and $s_v'$ with $s_v \xrightarrow{\mathit{bl}}_{\Van} s_v'$,
  \begin{enumerate}
  \item $\mathit{act}(\mathit{bl}) = \mathit{ra}\ q\ (\mathit{act}(s))\ \wedge\ \mathit{obs}(s) = \mathit{ob}\ q\ (\mathit{act}(s))\ h\ (\mathit{obs}(\mathit{bl})) \ \wedge\ \mathit{sec}(\mathit{bl}) = \mathit{sec}(s)$
    and 
  \item $\mathit{Sim}\ (\mathit{nxt}\ q\ (\mathit{act}(s)))\ (h \mathbin{+\!\!+} \mathit{obs}(\mathit{bl}))\ s'\ s_v'$; 
  \end{enumerate}
\item if $\mathit{Sim}\ q\ h\ s\ s_v$ and $s \in F_{\Opt}$, then there
  are $\mathit{bl}$ and $s_v' \in F_{\Van}$ with
  $s_v \xrightarrow{\mathit{bl}}_{\Van} s_v'$,
  $\mathit{act}(\mathit{bl}) = \mathit{fa}\ q$ and
  $\mathit{sec}(\mathit{bl}) = []$.
\end{enumerate}
Then $\Opt$ is relatively secure with respect to $\Van$ at the step
level.
\end{theorem}

Because the reference's actions are a function ($\mathit{ra}$, $\mathit{fa}$, $\mathit{nxt}$) of the LLM agent's actions only, two LLM-agent runs with equal actions are matched by two reference runs with equal actions, and because the LLM agent's observations are a function ($\mathit{ob}$) of the reference's, two LLM-agent runs that are observed differently are matched by reference runs that are observed differently. A block may be empty, so that a reference run lags, or long, so that it runs ahead. Block simulation proves \exAuto{}, \exMeeting{}, \exConfirm{}, \exSecond{} and \exStatus{}. In \exMeeting{}, a control state counts the steps and records the email that the LLM agent has read. When the LLM agent reads the injection, the reference's block reads a meeting request instead. The injected send is then matched by the reference's next two steps (its branch and its planned send), the LLM agent's own branch by an empty block, and every other step by one step with the same action. The observation of a block is its last one, and the final block is empty.

\subsection{Stuttering simulation}\label{sec:stutter}

For the attack of \exCallApp{} at the event level no relation meets the conditions of the
existing unwinding theorem. In the two runs of its step-level leak, the
injected call goes out at once in the run where it equals the approved
one, while the other run is still at the approval, so the relation
would have to hold between an interaction point and a state that is
not one, which those conditions forbid. The composites are deterministic, so write
$\mathit{next}_X(s)$ for the successor of a non-final state $s$ of
system $X$. The \emph{event contribution} $\mathit{ev}(s)$ of a
composite state is the pair of its event-level interaction
(\cref{def:event}) and the secret it produces, each $\bot$ if there is
none.

\begin{theorem}[Stuttering simulation]\label{thm:stutter}
Let $\Opt$ and $\Van$ be composites over the same configurations, with
$I_{\Opt} = I_{\Van}$, let $\mathit{Sim} \subseteq S_{\Opt} \times S_{\Van}$, let
$\mathit{Inv} \subseteq S_{\Opt}$, and let $w : S_{\Opt} \times S_{\Van} \to W$
for a well-founded order $(W, <)$. Suppose that
\begin{enumerate}
\item every $s \in I_{\Opt}$ satisfies $\mathit{Sim}\ s\ s$, $\mathit{Inv}\ s$
  and $\mathit{corr}(s, s)$;
\item if $\mathit{Sim}\ s\ s_v$ and $\mathit{Inv}\ s$, then $s \notin F_{\Opt}$,
  $s_v \notin F_{\Van}$ and $\mathit{Inv}\ (\mathit{next}_{\Opt}(s))$;
\item if $\mathit{Sim}\ s\ s_v$ and $\mathit{Inv}\ s$, then either
  \begin{enumerate}
  \item $\mathit{ev}(s) = (\bot, \bot)$,
    $\mathit{Sim}\ (\mathit{next}_{\Opt}(s))\ s_v$ and
    $w(\mathit{next}_{\Opt}(s), s_v) < w(s, s_v)$, or
  \item $\mathit{ev}(s) = \mathit{ev}(s_v)$ and
    $\mathit{Sim}\ (\mathit{next}_{\Opt}(s))\ (\mathit{next}_{\Van}(s_v))$.
  \end{enumerate}
\end{enumerate}
Then $\Opt$ is relatively secure with respect to $\Van$ at the event
level on possibly infinite runs.
\end{theorem}

Condition~2 restricts the theorem to systems whose runs never end, such as loops. The measure rules out that the reference stutters for ever. As in \cref{thm:compose}, each LLM-agent run is matched by one reference run, the reference run from the same initial state, which is why that state must correspond to itself. For the plan language it does, since the correspondence of \cref{sec:rs} is reflexive.

\myparagraph{Uniform verdicts.}\label{sec:uniform} Composition, block
simulation and stuttering simulation also give uniform relative security
(\cref{def:ursecure}), because they match each LLM-agent run with one
reference run, rather than each pair of runs with a pair. We call this
\emph{uniform matching}: there are fixed functions $F$ and $G$ such that
every completed LLM-agent run $\tr$ has a completed reference run $\trv$
that starts in a corresponding state, reads the same secrets, takes the
actions $F(\Act(\tr))$, and determines the LLM agent's observations as
$\Obs(\tr) = G(\Act(\tr), \Obs(\trv))$. Two LLM-agent runs with equal
actions are then matched by reference runs with equal actions, and if
the LLM agent's observations differ, so do the reference's. Composition,
and hence the fence theorems, and stuttering simulation match each run
with the reference run on the same actions and observations, so $F$ and
$G$ are the identity. 

\subsection{Disproof}\label{sec:disproof}

Step-level leaks are proved by the existing secret-directed unwinding theorem~\cite{derrick2025jar,afp-sdunwind}. To refute relative security at this level it suffices to give two completed LLM-agent runs with equal actions and different observations, and a relation $\Gamma$ on pairs of reference states, each paired with the secrets it has still to produce, such that (i) $\Gamma$ holds for all reference initial states corresponding to those of the two runs, paired with the two runs' secrets, (ii) related states are final together, (iii) related non-final states with equal actions have equal observations, and (iv) $\Gamma$ is preserved by steps with equal actions that consume the secrets. Then no two reference runs with those secrets and equal actions have different observations, so the leak is the LLM agent's alone.

In the examples $\Gamma$ typically relates reference states at the same \plan{} position, since the \plan{}'s control flow does not depend on the secret and the \plan{} never sends it. In \exSum{} the \plan{} reads the secret, reads the email and computes a summary, so neither reference run is ever observed, while the LLM-agent runs, with account numbers $s_1 \neq s_2$ and the injected email, are observed calling $\Fetch\ [s_1]$ and $\Fetch\ [s_2]$. Each disproof is therefore also a concrete test: two secrets, one attacker input, and the observation that tells the runs apart. Event-level leaks follow from the step-level ones by \cref{thm:transfer} where both composites have alignments, or directly when the reference makes no request, as in \exSum{}.


%% file: sections/examples.tex
\section{Case Studies}\label{sec:examples}

\Cref{tab:verdicts} lists the verdicts for a set of litmus tests.
Further details of the case studies are provided in
\cref{app:examples,app:prog,app:dual}, with models of three reported attacks in \cref{app:github}. The cases are litmus tests for the definitions, not general implementations of the named defences. A secure verdict without a scope shows that the criterion accepts that attack, not that the
defence stops every injection. \exFenced{} and \exSession{} are proved
for their attacks, and their general forms are
\cref{thm:revoke,thm:revoke-inf}. \Cref{sec:informal}
discusses fence placement and \cref{sec:loops} loops. 

\begin{table}[t]
  \caption{Relative-security verdicts of the case studies. Each
    verdict holds at both the event and the step level unless the
    Verdict column names a level. A plain \emph{secure} covers only the
    case's own attack. A scope in brackets names the class of attacks a
    secure verdict is proved for, and $\infty$ marks a loop, judged on
    possibly infinite runs.}\label{tab:verdicts}
\scriptsize\setlength{\tabcolsep}{3pt}\renewcommand{\arraystretch}{0.92}
\begin{tabular}{@{}>{\raggedright\arraybackslash}p{0.2\linewidth}>{\raggedright\arraybackslash}p{0.3\linewidth}>{\raggedright\arraybackslash}p{0.16\linewidth}>{\raggedright\arraybackslash}p{0.15\linewidth}>{\raggedright\arraybackslash}p{0.13\linewidth}@{}}
\toprule
Case & Agent & Monitor & Verdict & Method \\ \midrule
\exFenced{} & as \exSum{} & capability, fence & secure & fence thm; unwinding \\
\exEarly{} & $\Lock$ before the untrusted read & lock & secure [all] & lockstep \\
\exAuto{} & replies with the calendar & capability & secure & block \\
\exMeeting{} & sends the calendar by design & capability & secure & block \\
\exSession{} & fresh session per iteration & session fence & secure$^\infty$ & fence thm; unwinding \\
\exConfirm{} & calendar once the request is repeated & capability & secure & block \\
\exSecond{} & calendar if the first email, or else the second, is a request & capability & secure & block \\
\exStatus{} & report posted publicly, injection sends it & capability & secure & block \\
\exAllow{} & injected $\Fetch$, allowed URL only & capability, contract & secure & lockstep \\
\exBudgetFirst{} & one \Send{} before the read & budget & secure [\Send{}] & lockstep \\
\midrule
\exSum{} & injected $\Fetch$ carries the secret & capability & leaks & SD \\
\exLate{} & $\Lock$ after the secret & lock & leaks & SD \\
\exMiddle{} & $\Lock$ after the read & lock & leaks memory & SD \\
\exReset{} & injection adds the reset code & capability & leaks & SD \\
\exStale{} & \Send{} re-granted, old email in context & capability, fence, re-grant & leaks$^\infty$ & SD \\
\exToolApp{} & injected \Send{} after an approval & tool approval & leaks$^\infty$ & SD \\
\exUntagged{} & dual LLM without tags & allow all & leaks & SD \\
\exBranch{} & \plan{} branches on a tagged answer & tag check & leaks & SD \\
\exBudgetRace{} & injection spends the budget first & budget & leaks & SD \\
\midrule
\exCallApp{} & injected \Send{} matches the approval & call approval & secure$^\infty$ [\Send{}, fair] (event); leaks (step) & stutter; SD \\
\exTagged{} & dual LLM with CaMeL-style tags & tag check & secure [all]; leaks if requests shown & unwinding; SD \\
\exProbe{} & injected call on an invisible tool & capability, grant & secure [all]; leaks if denials seen & unobserved; SD \\
\exPoison{} & injected program overwrites a variable & capability, fence & leaks (persist); secure [no call or read] or in a sandbox (event) & SD; lockstep, sandbox thm \\
\exCondFetch{} & program reads a page chosen by the secret & capability, fence & secure [no call]; leaks if reads seen & unobserved; SD \\
\exSpot{} & spotlighting & capability & leaks; secure if the \model{} honours the marking & reduction; lockstep \\
\bottomrule
\end{tabular}

\smallskip
\begin{minipage}{\linewidth}\scriptsize
  \textbf{Scopes.} An attack is a parser and an oracle. [all]: every attack. [\Send{}]: every attack whose instructions are \Send{} calls. [\Send{}, fair]: such attacks in which the \model{} also refuses infinitely often. [no call or read], [no call]: every attack whose injected programs make no call or read, or no call.\\
  \textbf{Methods.} fence thm: \cref{thm:revoke,thm:revoke-inf}. sandbox thm: \cref{thm:sandbox}. unwinding: the existing unwinding theorem~\cite{derrick2025jar,afp-relsec}. lockstep, block, stutter: \cref{thm:lockstep,thm:block,thm:stutter}. SD: secret-directed unwinding (\cref{sec:disproof}). reduction: to \exSum{}. unobserved: \attacker{} sees nothing. \textbf{Other.} Monitors
  are defined in \cref{ex:comp-with-monit} and \cref{app:monitors}.
\end{minipage}
\vspace{-10pt}
\end{table}

\myparagraph{Legitimate disclosure.} \exAuto{}, \exMeeting{}
(\cref{ex:meeting-reply}), \exConfirm{}, \exSecond{} and \exStatus{}
disclose the user's data by design. \exAuto{} replies to every email
with the calendar and 
\exConfirm{} only sends the calendar once \attacker{} repeats its
request.  \exSecond{} gives \attacker{} a second chance, reading a
second email when the first is not a meeting request and sending the
calendar if the second one is.

In \exSecond{} the reference run that reproduces the injected leak is
shorter, since a meeting request as the first email makes it send the
calendar at once and stop. The existing unwinding theorem keeps the
runs in step and does not apply directly, so block simulation proves it
(\cref{sec:block}).

\exStatus{} shows that the leak need not use the same tool. On an empty
email, the \plan{} posts a secret report on a public status page, and
the injection makes the LLM agent send the same report through \Send{}
instead. Relative security accepts this, since it compares which
reports \attacker{} can tell apart, not which tool reveals them. 

Relative security 
rejects \exReset{} (\cref{ex:reset-code}). 
At both levels, the reference of \exReset{} discloses at most the
calendar (\cref{def:discloses}), but the LLM agent breaks this bound.

\myparagraph{Budgets and approvals.} A budget monitor allows a fixed
number of \Send{} calls per run. In \exBudgetRace{}, with a budget of
one, the \model{} obeys the injection before the \plan{}'s own
reply. The injected call sends the reset code and spends the budget,
so the planned reply with the calendar is denied, and \attacker{}
learns a secret that no reference run sends. In \exBudgetFirst{} the
\plan{} sends the calendar first, so the budget is spent before the
untrusted read and every injected \Send{} call is denied. 

Approving a tool is a budget of one that the injection can spend first
(\exToolApp{}, \cref{sec:approving-tool-or}). Approving the exact call
is secure at the event level for every parser of \Send{} calls and
every fair oracle, one that cannot obey for ever (\cref{sec:loops}),
and, unlike \exTagged{}, which refuses even the honest reply, it still
delivers the user's replies (\exCallApp{}). For such an oracle, every run's event-level observations are, for each approval $v$ in
turn, an email read and exactly $\Send\ [v]$, possibly sent by an
injected copy. So every approved reply is eventually sent and nothing
else is, and runs with the same approvals look alike. The proof needs
only a simulation relation, a measure that fairness makes finite, and
a one-step lemma, which is combined with \cref{thm:stutter} to achieve
the overall result.

\myparagraph{Heuristic defences and sanitisers.} Spotlighting marks
untrusted content and asks the \model{} not to obey it. We model the
marking as any injective function $\mathit{mark}$ applied to every
untrusted item. Since \attacker{} chooses the parser (\cref{sec:threat}),
it can choose $\mathit{mark}^{-1} \mathbin{;} \mathit{parse}$, where
$\mathit{parse}$ is the parser of \exSum{} (\cref{ex:sum-plan}), which
reads $\mathit{mark}(\vemail)$ as the same \Fetch{} call with the
secret. The marking therefore changes nothing, and \exSpot{} leaks for
every marking.\looseness=-1

So spotlighting works only if we assume that the \model{} honours the
marking. However, because the semantics leaves the \model{}'s choices
to \attacker{} it cannot establish that the verdict holds for a real
\model{}.
\label{sec:sanitiser}A trusted sanitiser $\sigma$ (\cref{fig:arch})
moves this assumption rather than removing it. It rewrites untrusted
content before the content enters the memory and the context, and the
theorem assumes $\mathit{parse}(\sigma(v)) = \bot$ for every $v$. This
\emph{clean-context} condition is on the sanitiser and the parser
together, since a parser that reads every value as an instruction
defeats every sanitiser. An implementation must establish it, for
instance with a typed interface that passes only values of a fixed
schema, as CaMeL's quarantined \model{} returns~\cite{debenedetti2025camel},
or, without proof, with a \model{} trained not to follow instructions
in data~\cite{chen2025struq,chen2025secalign}. Under the clean-context
condition the \model{}'s
context never holds an instruction, so the LLM agent takes exactly the
reference's steps. Relative security then holds for every \plan{} and
monitor, at both levels and on possibly infinite runs. 

\myparagraph{Dual-LLM agents.} In the dual-LLM pattern, the planner
never sees untrusted content, and a quarantined \model{} answers
queries on it. 
With CaMeL-style tags~\cite{debenedetti2025camel} and a monitor that
refuses a visible \Send{} call carrying a private value, \exTagged{}
is secure for every behaviour of the quarantined \model{}, at the
price of also refusing the honest reply. No \Send{} call reaches
\attacker{} in either system, so the verdict amounts to
noninterference. However, if \attacker{} sees refused requests,
\exTagged{} leaks if the refusal shows the request and the quarantined
\model{}'s answer depends on the notes. Without tags, the quarantined
\model{} can quote the user's notes into an answer the \plan{} then
sends (thus \exUntagged{} leaks).

In \exBranch{} the \plan{} branches on the quarantined answer and
sends \attacker{} a public constant, 1 or 0, in either branch. Every
\Send{} call passes the tag check, but an injected answer that depends
on the notes chooses the branch, and relative security fails at both
the step and event level. Our tags model CaMeL's NORMAL mode. Its
STRICT mode also tags values with the conditions that control them and
is designed to close this channel, which we consider future work.

\myparagraph{Contracts and what \attacker{} sees.} In \exAllow{}, a
contract on \Fetch{} lets an injected call through only if its
arguments are on an allow-list, here the single URL of a newsletter.
An allowed call therefore carries no secret. The case study models two
injected emails. One asks the agent to send the secret to
\attacker{}'s URL, which the contract denies, and the other asks it to
fetch the newsletter, which the contract allows. Against these two
attacks \exAllow{} is secure. However, other injections can still
leak. For example, one that fetches the newsletter if the secret is~0,
and \attacker{}'s URL otherwise, makes an allowed call for one secret
and a denied call for the other.  A step-level attacker sees whether
the call went out, so it learns whether the secret is~0, and relative
security fails.

The attacker models matter in \exProbe{}, where an injection calls an
internal tool that the user's setting, a secret, grants or not. No
tool is visible, so the event- and step-level attackers learn
nothing. However, if \attacker{} sees the tool on which a request is
refused (\cref{def:views}), for example through an error message, it
learns the setting, which the \plan{} never reveals. A defence is
secure only for a stated attacker model.\looseness=-1 

\myparagraph{Discussion.} As a practical takeaway, the verdicts support six rules for designing a monitor, each under the assumptions of the theorem or case behind it, namely to
\begin{enumerate*}[label=\bfseries(\arabic*)]
\item fence at the first untrusted read,
\item govern reads as well as calls,
\item approve exact calls rather than tools, which is secure at the event level only,
\item spend budgets before untrusted content arrives,
\item re-grant only with a fresh session, and
\item keep refusals from revealing arguments.
\end{enumerate*}

Relative security is neither monotonic nor compositional. If the
\plan{} sends a secret to \attacker{} by design and an injection fetches
a URL carrying it, the agent is secure while \Send{} is granted, but
revoking \Send{} removes the reference's matching leak (cf.\ \exStatus{}).
And \exSum{} is secure with only \Inbox{} and with only \Fetch{}, but
leaks with both.


%% file: sections/related.tex
\section{Related Work}\label{sec:related}

\myparagraph{Prompt injection.} Indirect prompt injection was
demonstrated against deployed LLM-integrated applications, including
exfiltration through tool
calls~\cite{greshake2023}. \citet{liu2024formalizing} formalise the
attack on a single query, without agents or properties over
runs. InjecAgent~\cite{zhan2024injecagent} and
AgentDojo~\cite{debenedetti2024agentdojo} measure how often agents
follow injected instructions, and adaptive attacks bypass twelve
defences, most of which had reported near-zero attack
success~\cite{nasr2025adaptive}. Our verdicts are instead about a
worst-case \model{}, within the threat model of \cref{sec:threat}. 

\myparagraph{Defences.} A number of works have proposed model-level
defences~\cite{chen2025struq,chen2025secalign}, which we state as
assumptions on the parser
(\cref{lem:ref}). Spotlighting~\cite{hines2024spotlighting} is
insecure under a worst-case \model{} (\exSpot{}). The dual-LLM
pattern~\cite{beurerkellner2025patterns} hands untrusted content to a
quarantined \model{} without tools, so an injection can corrupt an
answer but cannot add a request. It is one of our injection models,
and without tags it still leaks through a corrupted answer
(\exUntagged{}). CaMeL~\cite{debenedetti2025camel} adds an interpreter
that tags values with capabilities and checks policies before each
call, with security argued by design rather than proved.  \exTagged{}
models its tag check for a fixed \plan{} and is secure only because
every reply is refused, and \exBranch{} shows that tags that follow
data but not control leak. 

FIDES~\cite{costa2025fides} labels messages and tool results and lets
a call proceed only if its decision rests on trusted input, and
Conseca~\cite{tsai2025conseca} checks each call against a per-task
policy. The f-secure system~\cite{wu2024fsecure} keeps untrusted input
from the planner by a security monitor, with a formal integrity
guarantee, and Progent~\cite{shi2025progent} enforces privilege
policies over tool names and arguments. The fence of \exFenced{} is
FIDES's trusted-action policy with the whole context as the label, and
contracts model Conseca's policies (\exAllow{}). These systems enforce
\emph{absolute} policies, which neither imply nor follow from relative
security (\cref{tab:absolute}, \cref{app:absolute}).

\citet{siu2026formalizing} state agent security as properties of one
execution, such as action alignment and data isolation, against ideal
oracles that, for instance, identify which inputs acted as
instructions. Relative security instead relates the agent to a
reference that never obeys, so it does not require runtime instruction
attribution. AttriGuard~\cite{he2026attriguard} replays the agent on
weakened observations at run time and blocks a call that does not
recur. This attributes one call to untrusted content, empirically,
rather than bounding what \attacker{} can
distinguish. 

\myparagraph{Relative security.} Relative security was introduced for
speculative execution~\cite{dongol2024csf}, with a mechanisation in
Isabelle/HOL and unwinding methods for proof and
disproof~\cite{derrick2025jar,afp-relsec,afp-sdunwind}. It differs
from relative secrecy~\cite{volpano2000relative}, which bounds what a
program reveals by what match queries on the secret
reveal. Speculative semantics resolve the hardware's nondeterminism by
attacker-supplied directives~\cite{cauligi2020spectre} or a quantified
oracle~\cite{dongol2024csf}, as our parser and oracle do for the
\model{}. Speculative noninterference~\cite{guarnieri2020spectector} and
hardware-software contracts~\cite{guarnieri2021contracts} also compare
with a reference, but on the same inputs, whereas relative security
lets the reference use another input (\exMeeting{}).

Our novelties are the agent-and-monitor composition, the attacker
models and transfer, the fence theorems, lockstep for agents, and block
and stuttering simulation where the existing unwinding theorem is
incomplete. Block simulation is a strategy for the
existential runs, as in the verification of
hyperliveness~\cite{coenen2019hyperliveness} and of $\forall\exists$
properties of programs~\cite{beutner2022beyond,dickerson2022rhle}, and
stuttering simulation bounds the reference's stuttering by a measure,
as in stuttering bisimulation~\cite{namjoshi1997stuttering} and
CompCert~\cite{leroy2009compcert}.

\myparagraph{Information flow and declassification.}\label{sec:absolute}
An \emph{absolute} information-flow property judges the agent alone,
against a release policy. Noninterference releases nothing, and
declassification~\cite{sabelfeld2009declass} takes a policy written for
each task, whereas relative security takes it from the reference. 
Observational determinism~\cite{zdancewic2003observational} coincides
with noninterference here, since our semantics is
deterministic. Delimited release~\cite{sabelfeld2004delimited} agrees
with relative security on these agents under the task's policy, and
the disclosure bound is delimited release checked on the
reference. Robust
declassification~\cite{zdancewic2001robust,myers2004enforcing}, the
closest classical notion, forbids \attacker{} to steer what is
released, so it rejects \exMeeting{}'s own \plan{} unless its test is
endorsed. Explicit secrecy~\cite{schoepe2016explicit}, which taint
tracking and FIDES enforce, ignores implicit flows, so it accepts
\exBranch{}, every call of which passes the tag check. Integrity
counterparts such as transparent
endorsement~\cite{cecchetti2017nonmalleable} concern hijacking, and
the unwinding of relative security descends from that of
noninterference~\cite{goguen1984unwinding}.


%% file: sections/conclusion.tex
\section{Conclusion}\label{sec:conclusion}

\plant{} states prompt-injection security relative to a reference agent, which accepts useful disclosure and rejects injected leaks the reference cannot reproduce. Revoking the injectable tools at the first untrusted read, and never granting them back, suffices at the event level for every \plan{} and every \model{} behaviour of one-call injections on a schedule fixed in advance, on finite runs and, for a fair \model{}, on possibly infinite ones. A \model{} that decides when to obey by inspecting a secret or a monitor response, a tool whose reply depends on the call, and an untrusted planner are outside this guarantee. Mechanised verdicts show where other defences succeed and fail, and the disclosure bound makes explicit the leaks that the \plan{} makes by itself.

\myparagraph{Use of AI.} The Isabelle theorems and proofs were
generated with the help of Claude Code (Opus~5.5 and
Fable~5.1)~\cite{anthropic2025claudecode} with
PIDE-MCP~\cite{DBLP:journals/corr/abs-2604-15713,kappelmann2026pidemcp},
many inspired by the AFP entries~\cite{afp-relsec,afp-sdunwind}, and
Isabelle checked every proof. The authors checked each definition and
theorem of the paper against its Isabelle counterpart and take full
responsibility for any errors.




%% file: sections/appendix.tex
\section{The Remaining Case Studies}\label{app:examples}

This appendix describes the case studies of \cref{tab:verdicts} that
\cref{sec:informal} does not explain, except the injected programs
(\cref{app:prog}) and the dual-LLM agents (\cref{app:dual}). Unless stated otherwise, the secrets are trusted
inputs, the \model{} obeys \attacker{}'s email once, right after
reading it, and a verdict is for the example's own parser and oracle,
on finite runs.

\subsection{Monitors}\label{app:monitors}

Each monitor below is a triple $(\mu_0, \mathit{mdecide}, \mathit{mnext})$
(\cref{def:monitor}). In the examples that use the lock, the budget or
an approval, the only source tool is $\Inbox$, so a read is a request on
$\Inbox$ with arguments $[]$. In every monitor below a
denied request leaves the state unchanged, so a denial is silent to the
monitor as well. \Cref{def:monitor-ok} asks this only of requests in
$\Dcls$ made in deny mode. The general theorems hold for any
$\mathit{ok}$ and $\mathit{fail}$, so a \plan{} may check whether its
own call was allowed. The examples take $\mathit{ok} = \mathit{fail}$.
None of the example \plan{}s reads the
result of its own call, so the choice changes no verdict. It keeps the
two agents in step when the monitor denies a planned call in one and
allows it in the other, as in \exBudgetRace{}, where the injected
\Send{} spends the budget and the \plan{}'s own reply is denied. With
$\mathit{ok} \neq \mathit{fail}$ the two agents would then hold
different values in that call's variable, which the proofs would have to
track.

\myparagraph{Capability monitor and session fence} (\exSum{},
\exFenced{}, \exAllow{}, \exProbe{} and others). The state is a set
$\mathcal{C} \subseteq \mathit{Tool}$ of tools the agent may use, with
$\mu_0 = \mathcal{C}_0$. The parameters are a contract
$\mathit{contract} : \mathit{Tool} \to \mathit{Val}^{*} \to \mathit{Bool}$,
a grant $\mathit{grantOn} : \mathit{Val} \to 2^{\mathit{Tool}}$ for
trusted inputs, and the fenced set $\mathit{revokeOnU}$.
\begin{align*}
\mathit{mdecide}\ \mathcal{C}\ t\ \mathit{args} & \defeq
\begin{cases}
(\mathsf{Allowed}, \mathcal{C}) & \text{if } t \in \mathcal{C} \wedge \mathit{contract}\ t\ \mathit{args},\\
(\mathsf{Denied}, \mathcal{C}) & \text{otherwise,}
\end{cases}
\\
\mathit{mnext}\ \mathcal{C}\ \ell & \defeq
\begin{cases}
\mathcal{C} \cup \mathit{grantOn}\ v & \text{if } \ell = \mathsf{AReadT}\ v,\\
\mathcal{C} \setminus \mathit{revokeOnU} & \text{if } \ell = \mathsf{AReadU}\ t\ v,\\
\mathcal{C} & \text{otherwise.}
\end{cases}
\end{align*}
The decision never changes the state. The fence is the second case of
$\mathit{mnext}$, which the composite applies only to an allowed read
(\cref{fig:compose}). The \emph{session fence} (\exSession{}) has a further parameter
$\mathit{grantNew} \subseteq \mathit{Tool}$ and adds the
case $\mathit{mnext}\ \mathcal{C}\ \mathsf{ANew} \defeq \mathcal{C} \cup
\mathit{grantNew}$, so a fresh \model{} session gives the fenced tools
back. A contract other than $\lambda t\ \mathit{args}.\ \mathit{true}$
is an allow-list on arguments (\exAllow{}).

\myparagraph{Lock} (\exLate{}, \exMiddle{}, \exEarly{}). The state is a
flag, with $\mu_0 = \mathit{false}$.
\[
\mathit{mdecide}\ \mu\ t\ \mathit{args} \defeq
\begin{cases}
(\mathsf{Allowed}, \mathit{true}) & \text{if } t = \Lock,\\
(\mathsf{Allowed}, \mu) & \text{if } t = \Inbox \text{ or } \neg\mu,\\
(\mathsf{Denied}, \mu) & \text{otherwise,}
\end{cases}
\qquad
\mathit{mnext}\ \mu\ \ell \defeq \mu.
\]
The \plan{} raises the fence itself, by calling $\Lock$, so where the
fence goes is a property of the \plan{}. Reads stay allowed after the
lock, and every other tool is denied.

\myparagraph{Budget} (\exBudgetFirst{}, \exBudgetRace{}). The state is
a pair $(\mathcal{C}, \beta)$ of a capability set and a count
$\beta : \mathit{Tool} \to \mathbb{N}$ for each tool in a fixed set
$B$ of budgeted tools. With $\beta' = \beta[t \mapsto \beta\ t - 1]$
if $t \in B$ and $\beta' = \beta$ otherwise,
\begin{align*}
\mathit{mdecide}\ (\mathcal{C}, \beta)\ t\ \mathit{args} & \defeq
\begin{cases}
(\mathsf{Allowed}, (\mathcal{C}, \beta')) & \text{if } t \in \mathcal{C} \wedge \mathit{contract}\ t\ \mathit{args} \wedge (t \in B \Longrightarrow \beta\ t > 0),\\
(\mathsf{Denied}, (\mathcal{C}, \beta)) & \text{otherwise,}
\end{cases}
\\
\mathit{mnext}\ (\mathcal{C}, \beta)\ \ell & \defeq (\mathit{mnext}_{\mathit{cap}}\ \mathcal{C}\ \ell,\ \beta),
\end{align*}
where $\mathit{mnext}_{\mathit{cap}}$ is the capability monitor's
update. Unlike the capability monitor, a decision changes the state,
since every allowed request on a budgeted tool, planned or injected,
spends one unit. A budget therefore bounds how many calls happen, not
which ones, and whoever calls first wins (\exBudgetRace{}).

\myparagraph{Tool approval} (\exToolApp{}). The state is a flag $a$,
the user's approval, with $\mu_0 = \mathit{false}$.
\[
\mathit{mdecide}\ a\ t\ \mathit{args} \defeq
\begin{cases}
(\mathsf{Allowed}, a) & \text{if } t = \Inbox,\\
(\mathsf{Allowed}, \mathit{false}) & \text{if } a,\\
(\mathsf{Denied}, a) & \text{otherwise,}
\end{cases}
\qquad
\mathit{mnext}\ a\ \ell \defeq
\begin{cases}
\mathit{true} & \text{if } \ell = \mathsf{AReadT}\ v,\\
a & \text{otherwise.}
\end{cases}
\]
A trusted input approves the next \Send{} call, whatever its
arguments, and that call spends the approval.

\myparagraph{Call approval} (\exCallApp{}). The state is the approved
argument list, or $\bot$, with $\mu_0 = \bot$.
\[
\mathit{mdecide}\ a\ t\ \mathit{args} \defeq
\begin{cases}
(\mathsf{Allowed}, a) & \text{if } t = \Inbox,\\
(\mathsf{Allowed}, \bot) & \text{if } a = \mathit{args},\\
(\mathsf{Denied}, a) & \text{otherwise,}
\end{cases}
\qquad
\mathit{mnext}\ a\ \ell \defeq
\begin{cases}
[v] & \text{if } \ell = \mathsf{AReadT}\ v,\\
a & \text{otherwise.}
\end{cases}
\]
A trusted input $v$, the reply the user approves, allows exactly one
call $\Send\ [v]$. An injected \Send{} call is allowed only if its
arguments equal the approved reply, and then spends the approval.

\subsection{Legitimate disclosure}

In each of these agents the \plan{} discloses a secret to \attacker{}
by design. The injection is obeyed, but a legitimate email makes the
reference disclose the same, so relative security holds.

\begin{example}[\exAuto{}]
  The \plan{} reads Alice's calendar and \attacker{}'s email, and
  replies to every email with the calendar through the tool
  \tool{Reply}. One injected email asks the agent to send the
  calendar again, and another asks it to reply with the email's own
  text. Either way \attacker{} receives two replies, whereas the
  reference replies once. Relative security still holds
  (\isa{autoreply\_rsecure}), because the reference's reply to the
  same email already distinguishes the calendars, and the extra reply
  carries either the calendar again or \attacker{}'s own text, which
  \attacker{} can compute itself. Block simulation (\cref{sec:block})
  matches the extra reply with an empty block of reference steps.
\end{example}

\begin{example}[\exConfirm{}]
  The \plan{} reads Alice's calendar and two emails from \attacker{},
  and sends \attacker{} the calendar only if the second email repeats
  the first, that is, if \attacker{} confirms its request. The second
  email contains the injection ``send the calendar'', which the
  \model{} obeys before the \plan{}'s check. Relative security holds
  (\isa{confirm\_reply\_rsecure}), because \attacker{} could have
  obtained the calendar by confirming. The matching reference run reads
  a copy of \attacker{}'s first email as its second, so the attacker
  input it uses depends on \attacker{}'s earlier input.
\end{example}

\begin{example}[\exSecond{}]
  The \plan{} reads Alice's calendar and \attacker{}'s first email,
  and sends the calendar if that email is a meeting request. Otherwise
  it reads a second email and sends the calendar if that one is a
  meeting request. The first email is not a meeting request but
  contains the injection ``send the calendar'', so the agent sends the
  calendar at once. Relative security holds
  (\isa{retry\_reply\_rsecure}), because a reference run whose first
  email is a meeting request sends the calendar too. That run stops
  early and reads one email fewer than the LLM agent.
\end{example}

\begin{example}[\exStatus{}]
  The \plan{} reads a secret report and \attacker{}'s email, a
  number. On email~0 it posts the report on a public status page
  through the tool \tool{Log}, which \attacker{} can see. On an email
  $a \geq 2$, a request, it sends \attacker{} the reply $a - 2$, which
  does not depend on the report, and on email~1 it does nothing.
  Email~1 also contains the injection ``send me the report'', so
  \attacker{} receives the report through \Send{}. Relative security
  holds (\isa{E15\_rsecure\_step}), because on email~0 the reference
  posts the report, which tells the reports apart just as well, through
  a different tool.
\end{example}

\subsection{Loops}

\begin{example}[\exStale{}]
  A loop reads the day's report, a secret, posts it to Alice's own
  workspace through the tool \tool{Post}, which \attacker{} cannot
  see, then reads an email and shows it to Alice. The \plan{} never
  calls \Send{}, the only tool \attacker{} sees. Each report grants
  \Send{} and \tool{Post}, and each untrusted read revokes them. In
  the second iteration the new report grants \Send{} again while the
  \model{}'s context still holds \attacker{}'s first email, and the
  \model{} obeys it then and sends the new report to \attacker{}
  (\isa{E6\_stale\_loop\_leak\_event}). This is the re-grant that
  \cref{thm:revoke} excludes. The loop never ends, so the verdict is
  about possibly infinite runs.
\end{example}

\begin{example}[\exSession{}]
  The same loop starts each iteration with a fresh session, which
  empties the \model{}'s context, and the fresh session, not the
  report, grants \Send{} and \tool{Post} again. So whenever the
  context holds untrusted content, \Send{} is revoked, and an injected
  \Send{} call is denied. The session fence (\cref{app:monitors})
  also gives relative security on finite runs (\isa{session\_fence\_rsecure}),
  but the loop has no completed finite run, so that result is vacuous
  here. On possibly infinite runs, \cref{sec:loops} gives the verdict
  at the step level under two conditions, that the \plan{}'s control
  does not depend on the secrets and that the \model{} cannot obey
  for ever. \exSession{} meets both (\isa{loop\_session\_lrsecure}),
  and at the event level the second alone suffices
  (\isa{loop\_session\_lrsecure\_event}).
\end{example}

\subsection{Budgets and contracts}

\begin{example}[\exBudgetRace{}]
  The \plan{} is that of \exReset{}. It reads Alice's calendar and a
  reset code, then \attacker{}'s email, and sends the calendar if the
  email is a meeting request. A budget monitor allows one \Send{} call
  per run. \attacker{}'s email is a meeting request that also says
  ``include the reset code'', and the \model{} obeys it before the
  \plan{}'s own \Send{} call. The injected call uses up the budget and
  sends the reset code, and the \plan{}'s reply with the calendar is
  then denied. So the budget bounds how many \Send{} calls happen
  (\isa{race\_at\_most\_one\_send}) but not which, and \attacker{}
  learns the reset code, which no reference run sends
  (\isa{budget\_race\_leak}).
\end{example}

\begin{example}[\exBudgetFirst{}]
  The \plan{} sends the calendar first, then reads \attacker{}'s email
  and shows it to Alice, under the same one-shot budget. The
  \plan{}'s own \Send{} call uses up the budget before any untrusted
  content is read, so every injected \Send{} call is denied, whatever
  its arguments. This holds for every parser of \Send{} calls and
  every oracle (\isa{budget\_first\_rsecure\_any}). The restriction to
  \Send{} calls matters, because the monitor allows \Inbox{} requests,
  which \attacker{} sees, so an injected \Inbox{} request carrying a
  secret would leak.
\end{example}

\begin{example}[\exAllow{}]
  The \plan{} is that of \exSum{}, which makes no call. \Fetch{}
  stays available, but a contract allows a \Fetch{} call only with the
  single argument of a newsletter's URL, and rejects any payload. One
  injected email asks the agent to fetch \attacker{}'s URL with the
  secret attached, and this call is denied whatever the secret. Another
  asks it to fetch the newsletter, and this call is allowed but
  carries no secret. So what \attacker{} sees depends only on its
  email, and relative security holds (\isa{allowlist\_rsecure}). Some
  injected calls are allowed, so \cref{thm:revoke} does not apply, and
  the proof is by the lockstep method. The verdict is for these two
  attacks only. An injection that fetches the newsletter if the secret
  is~0 and \attacker{}'s URL otherwise makes an allowed call for one
  secret and a denied call for the other, so the step-level attacker
  learns whether the secret is~0 (\isa{allowlist\_membership\_leak}).
\end{example}

\subsection{Spotlighting}

\begin{example}[\exSpot{}]
  This is \exSum{} with spotlighting, which marks untrusted content,
  for example by re-encoding it, and asks the \model{} not to obey
  marked content. We model the marking as an injective function that
  the agent applies to content before the parser sees it. Any
  injective marking can be undone, since for every reaction to raw
  content there is a \model{} that reacts to the marked content in the
  same way (\isa{spotlight\_undone}). So the worst-case \model{} treats
  marked content as the \model{} of \exSum{} treats raw content, and
  the agent leaks for every injective marking
  (\isa{E10\_spotlight\_leak}). A \model{} that obeys no marked
  content gives relative security (\isa{E10\_honoured\_secure}). The
  two agents differ only in the \model{}, so this verdict rests on an
  assumption about the \model{} that the semantics cannot check.
\end{example}

\subsection{What the attacker sees}

\begin{example}[\exProbe{}]
  The \plan{} reads Alice's setting, a secret, and \attacker{}'s email,
  and shows the email to Alice, making no call. Setting~1 grants the
  agent an internal tool \tool{Log}, which \attacker{} cannot see, and
  the injected email makes the \model{} call \tool{Log}. With
  setting~1 the call is allowed, and with setting~0 it is refused. No
  tool is visible to \attacker{}, so the event- and step-level
  attackers learn nothing, for every parser and oracle
  (\isa{E16\_rsecure\_any}). An attacker who sees which tool a request
  was refused on, for example through an error message or a security
  log, learns the setting, which the \plan{} never reveals
  (\isa{E16\_not\_rsecure\_denial}).
\end{example}

\subsection{Reported attacks}\label{app:github}

\begin{example}[The GitHub MCP exploit]
  In the GitHub MCP exploit~\cite{invariant2025github}, Alice asks an
  agent to look at the open issues of her public repository. The agent
  can also read her private repositories. An issue filed by
  \attacker{} tells it to collect information about the repository's
  author and to publish it in a pull request on the public repository,
  and the agent does so with private data. This is \exSum{} with a
  reported tool set. The \plan{} reads Alice's private data, a secret,
  then an issue from the source tool \tool{Issues}, and shows a
  summary to Alice, making no call. The injected issue makes the
  \model{} call \tool{CreatePR} with the private data, and
  \tool{CreatePR}, which writes to the public repository, is the only
  tool \attacker{} sees. With every tool held throughout, two runs with
  the same injected issue and different private data open different
  pull requests, and the reference makes no call, so relative security
  fails at the event level (\isa{gh\_not\_rsecure}). With
  \tool{CreatePR} revoked at the first untrusted read,
  \cref{thm:revoke} gives relative security at the event level for
  every \plan{} over these tools, every parser whose instructions call
  \tool{CreatePR} and every oracle (\isa{Github\_Fenced.fenced\_rsecure}),
  and in the run of the exploit the injected pull request is denied
  (\isa{gh\_fenced\_blocks}). The model simplifies the attack in two
  ways. The reported agent fetches the private data with calls that the
  injection asks for, while here the \plan{} has already read it and one
  injected call publishes it. And the \plan{} here shows a summary,
  whereas the reported task was to address the issues, which may itself
  open a pull request. The fence would deny that call too. The Supabase
  MCP leak~\cite{generalanalysis2025supabase}, in which a support ticket
  makes a coding agent with a privileged database key write secret tokens
  into the ticket's thread, has the same shape; we do not model it.
\end{example}

\begin{example}[SpAIware]
  In SpAIware~\cite{rehberger2024spaiware}, a document or page that
  Alice asks ChatGPT about makes it store an instruction in its
  long-term memory, which is loaded into every later conversation. In
  each later answer the instruction makes ChatGPT show an image whose
  URL carries Alice's messages, and fetching the image sends them to
  \attacker{}. We model it as \exSession{} with a memory. Each
  iteration starts a fresh session, which empties the \model{}'s
  context, and then loads the memory by an untrusted read from the
  source \tool{Memory}, which puts the memory's content back into the
  context. It then reads Alice's message, a secret, and a page from
  \tool{Web}, and shows Alice an answer, making no call. The fence as
  deployed grants \tool{Img}, the image fetch and the only tool
  \attacker{} sees, at every fresh session, and revokes it at a page
  read but not at the memory load, which it counts as part of the
  session. The memory write \tool{Remember} is never revoked. In the
  first session the page makes the \model{} call \tool{Remember} with
  the instruction. In the second, the instruction loaded from the
  memory makes it call \tool{Img} with Alice's new message, after the
  fresh session has granted \tool{Img} again. Relative security fails
  on possibly infinite runs, at the step level
  (\isa{sp\_not\_lrsecure\_step}) and at the event level
  (\isa{sp\_not\_lrsecure}).

  The assumption the attack breaks is that of fresh sessions
  (\cref{sec:threat}): what the \model{} retains from untrusted
  content outlives \NewSession{}. Formally, \NewSession{} still empties
  the context, so ``the context is non-empty'' is still a sound taint
  (\cref{def:taint-ok}, \isa{sp\_lang\_taint\_ok}). But a session,
  memory load included, ends with the injected instruction in the
  context and \tool{Img} held (\isa{sp\_fresh\_session\_tainted}). The
  memory load is an untrusted read after which \tool{Img} is still
  allowed, so no deny mode meets the monitor obligations
  (\cref{def:monitor-ok}, \isa{sp\_not\_monitor\_ok}). Treating the
  memory as untrusted content, so that its load revokes \tool{Img} and
  \tool{Remember} like any other untrusted read and only a fresh
  session grants them, is the session fence. \Cref{thm:revoke-inf}
  then gives relative security at the event level on possibly infinite
  runs for every \plan{} over these tools, every parser whose
  instructions call \tool{Img} or \tool{Remember} and every fair
  oracle (\isa{Spaiware\_Fenced.fenced\_lrsecure}), and the attack's
  oracle is fair (\isa{sp\_fenced\_lrsecure}).

  The model simplifies the attack in three ways. Tool results come
  from fixed input streams, so the memory load in the second session
  returns \attacker{}'s instruction because the stream holds it, not
  because the write stored it. The fenced verdict holds for every
  stream, so it covers whatever the memory holds, but the model cannot
  judge the other fix, denying memory writes after an untrusted read
  while trusting the memory, which needs the memory as state. The
  \model{} obeys the stored instruction in one later session, not in
  every one. And the exfiltration is a single image fetch.
\end{example}

\begin{example}[MCP tool poisoning and shadowing]
  In tool poisoning~\cite{invariant2025toolpoisoning}, a malicious MCP
  server hides an instruction in the description of one of its tools,
  which the client loads into the \model{}'s context when a session
  starts. The instruction tells the agent to pass Alice's SSH key to
  that tool in a hidden argument, so that the key reaches \attacker{}'s
  server. In shadowing, it tells the agent to send Alice's email
  through a trusted server's send tool to \attacker{}'s address. The
  threat model covers both without change. The descriptions are
  untrusted content, read by the \plan{} before anything else from the
  source tool \tool{Mal}, the malicious server, which is also a tool
  \attacker{} sees. Alice's request is a trusted input, and her key is
  in the agent's memory. A trusted mail server has one send tool per
  recipient, and \attacker{} sees only the mail to its own address.
  Without a monitor, relative security fails for both attacks at the
  event level. Under tool poisoning, two runs with different keys pass
  different keys to \tool{Mal}, while the reference makes no call
  (\isa{tp\_not\_rsecure}). Under shadowing, two runs with different
  messages send \attacker{} different mail, while the reference mails
  only Alice's recipient (\isa{sh\_not\_rsecure}), although the
  injected call is on the trusted server. Since the descriptions are
  read first, a fence that revokes, at the untrusted read, every tool
  but Alice's own send tool closes at the start of the session.
  \Cref{thm:revoke} then gives relative security at the event level for
  every \plan{} over these tools, every parser whose instructions do
  not mail Alice's recipient and every oracle
  (\isa{Mcp\_Fenced.fenced\_rsecure}). Under the fence the poisoned
  call is denied, and so is the shadowed mail, while Alice's own mail
  still goes out (\isa{tp\_fenced\_blocks}, \isa{sh\_fenced\_blocks}),
  because a trusted interpreter runs the \plan{}'s send. In the reported
  attack the \model{} plans and redirects that send itself, which trusted
  execution rules out, and the fence would deny it, so the mail would not
  arrive.
  A fence that revokes only the server that supplied the descriptions
  does not stop shadowing (\isa{sh\_scoped\_not\_rsecure}). The model
  simplifies the attacks in three ways. The key is in memory rather
  than read by an injected call. The injected calls are separate calls,
  where the real attacks ride on the \plan{}'s own call to the
  malicious tool or replace its mail. And the recipient is part of the
  tool's name.
\end{example}

\subsection{Absolute properties}\label{app:absolute}

\Cref{tab:absolute} compares the absolute properties of \cref{sec:absolute} with relative security on three of the case studies.

\begin{table}[h]
\centering
\caption{Absolute properties and relative security (RS). \cmark:
  accepted, \xmark: rejected. } \label{tab:absolute}
\footnotesize
  \setlength{\tabcolsep}{2pt}
\begin{tabular}{@{}lcccc@{}}
\toprule
 & Releases & \shortstack{\exname{Meeting}\\\exname{Reply}} & \shortstack{\exname{Reset}\\\exname{Code}} & \shortstack{\exname{Tagged}\\\exname{Branch}} \\ \midrule
Noninterference & nothing & \xmark & \xmark & \xmark \\
Delimited release & per task & \cmark$^a$ & \xmark$^b$ & \xmark$^b$ \\
Robust declass. & per task & \xmark$^c$ & \xmark & \xmark$^c$ \\
Explicit secrecy & nothing & \xmark & \xmark & \cmark$^d$ \\
RS & reference & \cmark & \xmark & \xmark \\
\bottomrule
\end{tabular}
\par\smallskip{\footnotesize
$^a$\,If the policy releases the calendar.
$^b$\,Under the task's policy, and accepted by a looser one.
$^c$\,Unless the test is endorsed.
$^d$\,The leak is an implicit flow.\par}
\end{table}


%% file: sections/prog.tex
\section{Program Injection}\label{app:prog}

In the injection model of \cref{sec:inject} an obeyed instruction is a
single call. In general it is a program. This appendix gives the model
in which the parser yields programs, which the LLM agent runs in frames
on top of the \plan{}. The one-call model is its instance with
one-command programs (\cref{prop:one-call}).

\myparagraph{Parameters.} A \emph{program model} is a tuple
$(\mathit{parse}_P, \mathit{pick}, \mathit{onext}, o_0, \mathit{persist})$
with
\[
\mathit{parse}_P : \mathit{Val} \to (\mathit{Com}^{*})_{\bot}
\qquad
\mathit{persist} \in \mathit{Bool},
\]
and $\mathit{pick}$, $\mathit{onext}$ and $o_0$ as in
\cref{def:model}. If $\mathit{parse}_P\ v = p$, the content $v$
contains a program $p$, a list of commands of \cref{sec:plan}. The flag
$\mathit{persist}$ fixes what happens to an injected program's writes.
With $\mathit{persist}$ they reach the memory of the program below it
when it ends, and without it, in a \emph{sandbox}, they are discarded.

\myparagraph{Frames.} A \emph{frame} $(p, k, \mathit{fm})$ is a program
$p$, its program counter $k$ and its own memory $\mathit{fm}$. A
configuration of the LLM agent is a triple $(\kappa, \Phi, J)$ of an agent
configuration $\kappa$ of \cref{sec:plan}, a stack $\Phi$ of frames, and a
stream $J \in \mathit{Val}^{\omega}$ of items that frames read,
chosen by \attacker{} like $U$. Initially $\Phi = []$, $\kappa$ is initial as in \cref{sec:plan},
and $J$ is arbitrary. The program the \model{} would start in
$\kappa = \langle \pc, \mem{}, T, U, C, o, \pi \rangle$ is
\[
  \mathit{injProg}\ \kappa \defeq
  \text{if } \mathit{pick}\ o = i \text{ and } i < |C| \text{ then }
  \mathit{parse}_P\ (C[i]) \text{ else } \bot .
\]

\myparagraph{Frame steps.} A frame runs one command of its program at
a time, on its own memory:
\begin{itemize}
\item $\Call\ x\ t\ g$ emits $\mathsf{ACall}\ t\ (g\,\mathit{fm})$ and does
  not store the result;
\item $\ReadU\ x\ t$ takes the next content $v$ of $J$, emits
  $\mathsf{AReadU}\ t\ v$, and leaves the read $(x, v)$ pending;
\item $\mathsf{Assign}\ x\ f$ and $\mathsf{IfJump}\ b\ n_1\ n_2$ act on
  $\mathit{fm}$ as in \cref{fig:sos};
\item every other command is skipped, so a frame reads no trusted input
  and starts no session.
\end{itemize}
Each frame step also advances the \model{}'s state to $\mathit{onext}\ o$.
A frame whose counter has passed the end of its program is popped in
the same step as its last command. With $\mathit{persist}$, its memory
then replaces that of the frame below it, or the \plan{}'s memory if
there is none. In a sandbox nothing else changes. Write
$\mathit{fstep}\ \mathit{fr}\ \Phi\ \kappa\ J$ for the label and the
configuration after one step of frame $\mathit{fr}$ on the stack $\Phi$.

\myparagraph{The LLM agent.} The LLM agent for programs has three rules.
\begin{mathpar}
\inferrule[P-Plan]
  {\mathit{injProg}\ \kappa = \bot \\ \kappa \xrightarrow{\ell}_{\Van} \kappa'}
  {(\kappa, [], J) \xrightarrow{\ell}_{P} (\kappa', [], J)}
\and
\inferrule[P-Start]
  {\mathit{injProg}\ \kappa = p \\ \Phi = [] \Longrightarrow (\pc < |\plansf| \wedge \pi = \emptyset)}
  {(\kappa, \Phi, J) \xrightarrow{\ell}_{P} \kappa_1
   \quad\text{where } (\ell, \kappa_1) = \mathit{fstep}\ (p, 0, \mathit{top}\ \Phi\ \kappa)\ \Phi\ \kappa\ J}
\and
\inferrule[P-Frame]
  {\mathit{injProg}\ \kappa = \bot}
  {(\kappa, \mathit{fr} \cdot \Phi, J) \xrightarrow{\ell}_{P} \kappa_1
   \quad\text{where } (\ell, \kappa_1) = \mathit{fstep}\ \mathit{fr}\ \Phi\ \kappa\ J}
\end{mathpar}
Here $\mathit{top}\ \Phi\ \kappa$ is the memory of the top frame of $\Phi$,
or the \plan{}'s memory $\mem{}$ if $\Phi = []$. So the \model{} may start
a program whenever it obeys a context item, while the \plan{} runs or
inside another injected program, and the new program runs on a copy of
the memory below it. Its first command runs in the same step. The
monitor's response to a frame's request is received by the top frame. An
allowed read stores the content in the frame's memory and appends it to
the context $C$, where it may be parsed in turn. A denied read stores
$\mathit{fail}$. A read that is the last command of its program has no
effect beyond its request, since its frame is popped before the response
arrives. The reference is the \plan{} alone, as in
\cref{sec:plan}, and leaves $\Phi$ and $J$ unchanged.

\begin{proposition}[One-call programs]\label[proposition]{prop:one-call}
For a parser $\mathit{parse}$ of \cref{def:model} and a variable $x_0$,
let $\mathit{parse}_P\ v \defeq [\Call\ x_0\ t\ g]$ if
$\mathit{parse}\ v = (t, g)$, and $\bot$ otherwise. For every \plan{},
oracle, value of $\mathit{persist}$ and monitor, the program model with
$\mathit{parse}_P$ is relatively secure at the event level iff the
one-call model with $\mathit{parse}$ is.
\end{proposition}

A one-command program that calls is popped in the step that makes the
call, so the frame stack is empty between steps and the frame stream is
never read. For each frame stream, runs of the two models then
correspond one to one, and are viewed alike. The proposition is stated for the
event level on finite runs, which is what is mechanised.

\myparagraph{The fence for programs.} \Cref{thm:revoke} is for one-call
injection. With $\mathit{persist}$ the fence is not enough. An injected
program that makes no call can still overwrite a variable the \plan{}
later sends (\exPoison{}). In a sandbox it is enough, provided the fence
covers reads as well as calls.

\begin{theorem}[Sandboxed programs]\label{thm:sandbox}
In the program model without $\mathit{persist}$, suppose that every call $\Call\ x\ t\ g$ in a program that $\mathit{parse}_P$ yields satisfies $\Dcls\ t\ (g\,\mathit{fm})$ for every memory $\mathit{fm}$, that every read $\ReadU\ x\ t$ in such a program satisfies $\Dcls\ t\ []$, and that $\deny$ and $\Dcls$ meet the monitor obligations. Then the composite is relatively secure at the event level, for every \plan{}, every such parser and every oracle, however deeply the \model{} nests injected programs.
\end{theorem}

By \cref{prop:one-call}, \cref{thm:revoke} is the case of one-command programs, which never read and leave no frame to sandbox.

\myparagraph{Examples.} The following two case studies use the program
model.

\begin{example}[\exPoison{}]
  Here an email may parse to a whole program, which runs in a frame on
  a copy of the \plan{}'s memory. The \plan{} reads
  a secret and \attacker{}'s email, and sends the email back to
  \attacker{}. A fence revokes \Fetch{} on the untrusted read, but
  \Send{} stays available because the \plan{} needs it. The injected
  program makes no call and only overwrites the variable that holds
  the email with the secret. If the frame's writes persist into the
  \plan{}'s memory, the \plan{}'s own \Send{} call then carries the
  secret, and the fence does not stop it, because the leaking call is
  the \plan{}'s own (\isa{E17\_persist\_not\_rsecure}). If the frame's
  writes are discarded, as in a sandbox, relative security holds
  (\isa{E17\_sandbox\_secure}), at the event level for every oracle and
  every parser of programs that neither call nor read
  (\isa{E17\_sandbox\_any}).
\end{example}

\begin{example}[\exCondFetch{}]
  The \plan{} reads a secret and \attacker{}'s email, and shows the
  email to Alice, making no call. The injected program, run in a
  sandboxed frame, reads one of two pages that \attacker{} serves, the
  first if the secret is~0 and the second otherwise. Reading a page is
  an untrusted read, not a call, so the program makes no call. While
  \attacker{} cannot see reads, relative security holds, for every
  parser of call-free programs and every oracle, sandboxed or not
  (\isa{E18\_rsecure}, \isa{E18\_rsecure\_any}). If \attacker{} runs the
  page servers and so sees which page is read, it learns the secret
  (\isa{E18\_not\_rsecure\_read}). A monitor that also denies reads
  from \attacker{}'s sources once untrusted content has been read
  prevents this (\isa{E18\_read\_fence\_rsecure}), and for agents that
  make no call and whose \plan{} reads none of \attacker{}'s sources
  this holds for every parser and oracle (\isa{read\_fence\_rsecure}).
\end{example}

%% file: sections/dual.tex
\section{Dual-LLM Agents}\label{app:dual}

The dual-LLM pattern~\cite{beurerkellner2025patterns}, on which CaMeL~\cite{debenedetti2025camel} builds,
splits the agent in two. A privileged planner holds the tools and runs
the \plan{}, and never sees untrusted content, so no instruction hidden
in that content can make it call a tool. A quarantined \model{} reads
the untrusted content, holds no tools, and only answers queries. An
injection can therefore add no call. It can only make an answer wrong,
and the \plan{} then uses that answer.

\myparagraph{Commands.} The dual-LLM language keeps the commands of
\cref{sec:plan} for trusted input, calls and jumps, replaces the
untrusted read, and adds a query:
\[
\begin{array}{r@{\;}c@{\;}l}
d & ::= & \kw{DReadT}\ x \mid \kw{DReadU}\ x\ t \mid \kw{DQuery}\ x\ \mathit{ys}
  \mid \kw{DCall}\ x\ t\ g \mid \kw{DJump}\ n \mid \kw{DIfJump}\ b\ n_1\ n_2
\end{array}
\]
with $\mathit{ys} \in \mathit{Var}^{*}$. $\kw{DQuery}\ x\ \mathit{ys}$
stores in $x$ the quarantined \model{}'s answer on the values of
$\mathit{ys}$. The \plan{} is fixed, so untrusted content read by
$\kw{DReadU}$ can affect a run only through the values that the \plan{}
computes with, among them the answers to queries. Configurations
are those of \cref{sec:plan}.

\myparagraph{Parameters.} The quarantined \model{} and the interpreter
are given by
\[
\begin{array}{r@{\;:\;}l@{\qquad}r@{\;:\;}l}
q_V & \mathit{Val}^{*} \to \mathit{Val} &
\mathit{tag}_T,\ \mathit{tag}_U & \mathit{Val} \to \mathit{Val} \\
q_O & \Omega \to \mathit{Val}^{*} \to \mathit{Val} &
\mathit{qtag} & \mathit{Val}^{*} \to \mathit{Val} \to \mathit{Val}
\end{array}
\]
together with $o_0$ and $\mathit{onext}$ as in \cref{sec:plan}. The
honest answer on inputs $\mathit{as}$ is $q_V\ \mathit{as}$, and the
answer in \model{} state $o$ is $q_O\ o\ \mathit{as}$, which sees
nothing but $o$ and $\mathit{as}$. Tags are part of the values, as in
CaMeL, and are set by the interpreter, not the \model{}.
$\mathit{tag}_T$ and $\mathit{tag}_U$ tag a trusted input and an
untrusted content when they are read, and $\mathit{qtag}\ \mathit{as}\ a$
tags an answer $a$ from its inputs $\mathit{as}$, whatever the \model{}
returned. With identity tag functions the agent is the plain dual-LLM
pattern.

\myparagraph{Steps.} For an answer function
$q : \mathit{Val}^{*} \to \mathit{Val}$, the step relation
$\xrightarrow{}_{q}$ has the rules
\begin{mathpar}
\inferrule[DReadT]
  {\plansf[\pc] = \kw{DReadT}\ x \\ T = v \cdot T'}
  {\langle \pc, \mem{}, T, U, C, o, \pi \rangle \xrightarrow{\mathsf{AReadT}\ v}_{q}
   \langle \pc{+}1, \mem{}[x \mapsto \mathit{tag}_T\ v], T', U, C, o', \pi \rangle}
\and
\inferrule[DReadU]
  {\plansf[\pc] = \kw{DReadU}\ x\ t \\ U = v \cdot U'}
  {\langle \pc, \mem{}, T, U, C, o, \emptyset \rangle \xrightarrow{\mathsf{AReadU}\ t\ v}_{q}
   \langle \pc{+}1, \mem{}, T, U', C, o', (x, \mathit{tag}_U\ v) \rangle}
\and
\inferrule[DQuery]
  {\plansf[\pc] = \kw{DQuery}\ x\ \mathit{ys} \\ \mathit{as} = [\mem{}\,y \mid y \leftarrow \mathit{ys}]}
  {\langle \pc, \mem{}, T, U, C, o, \pi \rangle \xrightarrow{\tau}_{q}
   \langle \pc{+}1, \mem{}[x \mapsto \mathit{qtag}\ \mathit{as}\ (q\ \mathit{as})], T, U, C, o', \pi \rangle}
\end{mathpar}
and the rules \textsc{Call} and \textsc{IfJump} of \cref{fig:sos} for
$\kw{DCall}$ and $\kw{DIfJump}$, with $\kw{DJump}\ n$ stepping as
$\kw{DIfJump}\ (\lambda \mem{}.\ \mathit{true})\ n\ n$, where again
$o' = \mathit{onext}\ o$. The labels carry the value read, not its
tagged form. The reference is ${\xrightarrow{}_{\Van}} \;=\; {\xrightarrow{}_{q_V}}$,
and the LLM agent steps from a configuration with \model{} state $o$ by
$\xrightarrow{}_{q_O\ o}$. The two agents thus run the same \plan{} and
differ only in the answers to queries. Neither makes an injected call,
so the parser and $\mathit{pick}$ (\cref{sec:plan}) play no part. Two
initial configurations correspond as in \cref{sec:rs}, and the agents
compose with a monitor by the rules of \cref{fig:compose}, so the
attacker models of \cref{sec:rs} and the transfer theorem apply
unchanged. A monitor
that decides on tags, as in \exTagged{}, reads them from the arguments
of a request.

The verdicts below are for one \plan{} each, and \exTagged{}'s holds
for every $q_O$.

\begin{example}[\exUntagged{}]
  The \plan{} reads
  Alice's private notes and \attacker{}'s email, asks the quarantined
  \model{} for a reply given both, and sends the reply to \attacker{}.
  The reference's quarantined \model{} answers honestly, and its reply
  depends only on the email. Under injection, the quarantined
  \model{} quotes the notes into the reply, which the \plan{} then
  sends (\isa{E9\_dual\_llm\_leak}). The split stops injected tool
  calls, since the quarantined \model{} makes none, but not injected
  content in a value that the \plan{} uses.
\end{example}

\begin{example}[\exTagged{}]
  The same \plan{} runs with CaMeL-style tags. The interpreter tags
  the notes as private and the email as public, and an answer of the
  quarantined \model{} is private if any of its inputs is. The monitor
  refuses a visible \Send{} call that carries a private value. The
  reply is computed from the notes, so it is always private, and the
  monitor refuses it in the LLM agent and in the reference alike. No
  \Send{} call ever reaches \attacker{} (\isa{e8\_obsO},
  \isa{e8\_obsV}), so relative security holds for every behaviour of
  the quarantined \model{} (\isa{camel\_rsecure}), and the verdict
  amounts to noninterference. The price is that the honest reply is
  refused too. An attacker who sees refused requests with their
  arguments learns about the notes whenever the quarantined \model{}'s
  answer depends on them (\isa{camel\_not\_rsecure\_denial\_request}),
  whereas showing only the refused tool is harmless
  (\isa{camel\_rsecure\_denial\_tool}).
\end{example}

\begin{example}[\exBranch{}]
  Tags that follow data but not control flow are not enough for every
  \plan{}. \exBranch{} keeps the tags and the monitor of \exTagged{},
  but its \plan{} branches on the quarantined answer and sends
  \attacker{} the constant~1 in one branch and~0 in the other. Both
  constants are public, so every \Send{} call passes the tag check
  (\isa{i\_label\_check\_passes}). An injected answer that is~1 exactly
  when the notes are~1 then chooses the branch, so \attacker{} learns
  whether the notes are~1. The reference, whose answer depends only on
  the email, never reveals this, and relative security fails at both
  levels (\isa{implicit\_flow\_leak}, \isa{implicit\_flow\_leak\_event}).
  CaMeL's STRICT mode, which also tags control dependencies, is
  designed to close this channel, and we do not model it.
\end{example}